\documentclass[useAMS,usenatbib]{mn2e}
\usepackage{graphicx}
\usepackage{lscape}

\title[The benchmark metal poor T8 dwarf BD+01 2920B]{Discovery of the benchmark metal poor T8 dwarf BD+01 2920B}
\author[D. J. Pinfield et~al.]{D. J. Pinfield$^{1}$\thanks{E-mail: D.J.Pinfield@herts.ac.uk}, 
B. Burningham$^{1}$, N. Lodieu$^{2,3}$, S. K. Leggett$^{4}$, C. G. Tinney$^{5}$, 
\newauthor
L. van Spaandonk$^{1}$, F. Marocco$^{1}$, R. Smart$^{6}$, J. Gomes$^{1}$, L. Smith$^{1}$, P. W. Lucas$^{1}$, 
\newauthor
A. C. Day-Jones$^{7}$,  D. N. Murray$^{1}$, A. C. Katsiyannis$^{8}$, S. Catalan$^{1}$, C. Cardoso$^{6,1,9}$, 
\newauthor
J. R. A. Clarke$^{1}$, S. Folkes$^{10}$, M. C. G\'{a}lvez-Ortiz$^{11}$, D. Homeier$^{12}$, J. S. Jenkins$^{7}$, 
\newauthor
H. R. A. Jones$^{1}$, Z. H. Zhang$^{1}$ \\
$^{1}$Centre for Astrophysics Research, Science and Technology Research
Institute, University of Hertfordshire, Hatfield AL10 9AB \\
$^{2}$Instituto de Astrof\'isica de Canarias, V\'{i}a L\'{a}ctea s/n, E-38205 La Laguna, Tenerife, Spain \\
$^{3}$Departamento de Astrof\'isica, Universidad de La Laguna (ULL), E-38205 La Laguna, Tenerife, Spain \\
$^{4}$Gemini Observatory, 670 N. A'ohoku Place, Hilo, HI 96720, USA \\
$^{5}$Department of Astrophysics, School of Physics, University of New South Wales, NSW 2052, Australia \\
$^{6}$Istituto Nazionale di Astrofisica, Osservatorio Astronomico di Torino, Strada Osservatrio 20, 10025 Pino Torinese, Italy \\
$^{7}$Universidad de Chile, Santiago, Casilla 36-D, Chile \\
$^{8}$Institute of Astronomy \& Astrophysics, National Observatory of Athens, Penteli, GR-15236, Greece \\
$^{9}$Astrophysics Group, College of Engineering, Mathematics, and Physical Sciences, University of Exeter, Exeter EX4 4QL, UK \\
$^{10}$Departamento de F\'isica y Astronom\'ia, Universidad de Valpara\'iso, Av. Gran Breta\~na 1111, Casilla 5030, Valpara\'iso, Chile \\
$^{11}$Centro de Astrobiolog\'ia (CSIC-INTA), Ctra. Ajalvir km 4, E-28850 Torrej\'on de Ardoz, Madrid, Spain\\
$^{12}$CRAL, UMR 5574, CNRS, Universit\'e de Lyon, \'Ecole Normale Sup\'erieure de Lyon, 46 All\'ee d'Italie, 69364 Lyon C\'{e}dex 07, France
}

\begin{document}
%
%
%
%


\def\aj{\rm{AJ}}                   
\def\araa{\rm{ARA\&A}}             
\def\apj{\rm{ApJ}}                 
\def\apjl{\rm{ApJ}}                
\def\apjs{\rm{ApJS}}               
\def\ao{\rm{Appl.~Opt.}}           
\def\apss{\rm{Ap\&SS}}             
\def\aap{\rm{A\&A}}                
\def\aapr{\rm{A\&A~Rev.}}          
\def\aaps{\rm{A\&AS}}              
\def\azh{\rm{AZh}}                 
\def\baas{\rm{BAAS}}               
\def\jrasc{\rm{JRASC}}             
\def\memras{\rm{MmRAS}}            
\def\mnras{\rm{MNRAS}}             
\def\pra{\rm{Phys.~Rev.~A}}        
\def\prb{\rm{Phys.~Rev.~B}}        
\def\prc{\rm{Phys.~Rev.~C}}        
\def\prd{\rm{Phys.~Rev.~D}}        
\def\pre{\rm{Phys.~Rev.~E}}        
\def\prl{\rm{Phys.~Rev.~Lett.}}    
\def\pasp{\rm{PASP}}               
\def\pasj{\rm{PASJ}}               
\def\qjras{\rm{QJRAS}}             
\def\skytel{\rm{S\&T}}             
\def\solphys{\rm{Sol.~Phys.}}      
\def\sovast{\rm{Soviet~Ast.}}      
\def\ssr{\rm{Space~Sci.~Rev.}}     
\def\zap{\rm{ZAp}}                 
\def\nat{\rm{Nature}}              
\def\iaucirc{\rm{IAU~Circ.}}       
\def\aplett{\rm{Astrophys.~Lett.}} 
\def\apspr{\rm{Astrophys.~Space~Phys.~Res.}}
\def\bain{\rm{Bull.~Astron.~Inst.~Netherlands}} 
\def\fcp{\rm{Fund.~Cosmic~Phys.}}  
\def\gca{\rm{Geochim.~Cosmochim.~Acta}}   
\def\grl{\rm{Geophys.~Res.~Lett.}} 
\def\jcp{\rm{J.~Chem.~Phys.}}      
\def\jgr{\rm{J.~Geophys.~Res.}}    
\def\jqsrt{\rm{J.~Quant.~Spec.~Radiat.~Transf.}}
\def\memsai{\rm{Mem.~Soc.~Astron.~Italiana}}
\def\nphysa{\rm{Nucl.~Phys.~A}}   
\def\physrep{\rm{Phys.~Rep.}}   
\def\physscr{\rm{Phys.~Scr}}   
\def\planss{\rm{Planet.~Space~Sci.}}   
\def\procspie{\rm{Proc.~SPIE}}   

\let\astap=\aap
\let\apjlett=\apjl
\let\apjsupp=\apjs
\let\applopt=\ao

\maketitle

\begin{abstract}
We have searched the WISE first data release for widely separated ($\le$10,000AU) late T dwarf companions to 
Hipparcos and Gliese stars. We have discovered a new binary system containing a K-band suppressed T8p dwarf 
WISEP J1423+0116 and the mildly metal poor ([Fe/H]=$-$0.38$\pm$0.06) primary BD+01 2920 (Hip 70319), a G1 dwarf at a 
distance of 17.2pc. This new benchmark has $T_{\rm eff}$=680$\pm$55K and a mass of $20-50 M_{Jup}$. Its spectral 
properties are well modelled except for known discrepancies in the $Y$ and $K$ bands. Based on the well determined 
metallicity of its companion, the properties of BD+01 2920B imply that the currently known T dwarfs are dominated 
by young low-mass objects. We also present an accurate proper motion for the T8.5 dwarf WISEP J075003.84+272544.8.
\end{abstract}


\section{Introduction}
\label{sec:intro}

An accurate understanding of the physics of ultra-cool atmospheres ($T_{\rm eff}<$ 2300K) is a major and 
ongoing challenge for theory \citep[e.g.][]{allard1997}. Complex molecular opacities \citep[e.g.][]{barber2006}, 
condensate clouds and their properties \citep[e.g.][]{allard2001}, and non-equilibrium chemistry \citep[i.e. 
vertical transport or mixing;][]{saumon2007} are significant sources of uncertainty in the models. However, it 
is crucial to improve our understanding if we are to effectively measure the properties of substellar 
populations (brown dwarfs and giant planets) and study their formation and evolution 
\citep[e.g.][]{bate2002,goodwin2007,stamatellos2007,sumi2011}.

Building on the samples of L ($\sim$2300K--1500K) and T dwarfs ($\sim$1500K--500K) \citep{kirkpatrick2005} 
identified in the Two-Micron All Sky Survey \citep{skrutskie2006}, the DENIS survey \citep{epchtein1997} and 
the Sloan Digital Sky Survey \citep{York2000}, a new generation of infrared surveys is expanding our search-space 
into much greater volumes. The UKIRT Infrared Deep Sky Survey \citep[UKIDSS;][]{lawrence2007} is sensitive to 
mid L -- mid T dwarfs out to $\sim$100pc over $\sim$15\% of the sky. The VISTA surveys (e.g. the Viking 
and VHS surveys) will expand this coverage to $\sim$50\% of the sky in the next few years. For the latest T dwarfs 
\citep[T8-9; e.g.][]{warren2007,ben2008,delorme2008,phil2010,ben2011b} the sensitivities of these surveys are 
matched by those of the Wide-field Infrared Survey Explorer \citep[WISE;][]{wright2010}, probing to distances 
of $\sim$15--25pc. And for even lower temperatures ($T_{\rm eff}<$500K) an increased mid-/near-infrared flux 
ratio is allowing WISE to uncover the new Y dwarf class in the $\sim$300--500K range \citep{cushing2011}. 
Together these surveys are characterising a rapidly growing population in the near- and mid-infrared 
\citep[e.g.][]{lodieu2007,me2008,ben2010b,reyle2010,kirkpatrick2011}.

With sensitivity to a growing field L-, T- and Y- dwarf search-space it is becoming feasible to search for multiple 
systems \citep[e.g.][]{ben2009,zenghua2010,ben2010a,avril2011,dm2011,ben2011a,sandy2010b} or moving group associations 
\citep[e.g.][]{clarke2010,maricruz2010}. The physical properties (mass, age and metallicity) of such objects 
can be constrained through association with more readily characterisable stellar companions or moving group 
members, establishing them as benchmark objects that can test the theory or more directly map physical 
properties onto spectral characteristics \citep[e.g.][]{me2006}. Indeed, by searching for even rarer benchmarks 
with better physical constraints which span a more extreme range of properties, it will be possible to provide 
the strongest tests for the model atmospheres, a goal that absolutely requires sensitivity to large volumes.

In this paper we present a search of the WISE first data release for widely separated late T dwarf companions 
to stars with known parallaxes. Section 2 describes our WISE sample selection, and Section 3 the method used to 
identify candidate binary associations. Sections 4 and 5 present our spectroscopic and additional photometric 
data, and in Section 6 we derive candidate proper motions. Section 7 statistically assesses the expected level 
of false positives in our search, and Sections 8, 9 and 10 discuss the properties and characteristics of a newly 
discovered benchmark system. Conclusions and future work are in Section 11.

\section{WISE sample}
\label{sec:sample}

We identified candidate mid-T and later type objects in the WISE Preliminary Data Release source catalogue, 
which we accessed via the NASA/IPAC Infrared Science Archive's catalogue query engine. We performed a 
series of all-sky searches using structured query language input to select sources with constraints on 
signal-to-noise and colour, and with detections in various combinations of bands chosen to optimise 
sensitivity to late T dwarfs. We always required a detection in the W2-band with signal-to-noise (SNR) 
of at least 10. If $W1$- and $W2$-band detections are available then we require $W1-W2\ge$2.0 to select spectral 
types of $>$T5 \citep{kirkpatrick2011,mainzer2011}. If $W2$- and $W3$-band detections are available 
then we require $W2-W3\le$2.5 in order to avoid dusty galaxies such as ULIRGS, LINERS and obscured AGN 
\citep{wright2010}. As well as our WISE-band detection requirements we also made use of the WISE catalogue 
cross-match with the 2MASS point source catalogue, to divide our searches into objects that are detected 
in 2MASS (within 3 arcseconds of the WISE position) and those that are not. For 2MASS detected objects we 
required that either $H-W2\ge$2.5 or $J-W2\ge$3.5 so as to remove L and early T dwarfs. The full complement 
of searches and the number of sources identified in each is shown in Table \ref{tab:search}.

The search requiring detection in only the $W2$-band will be the most sensitive to faint objects with red 
WISE colours since the WISE sensitivity limits \citep[all-sky 5-$\sigma$ Vega limits are $W1$=16.5, $W2$=15.5, 
$W3$=11.2, $W4$=7.9;][]{wright2010} mean that objects with $W1-W2>$2 will generally only be detected in $W2$ for 
$W2$=14.5-15.5 (i.e. at least $\sim$75\% of the $W2$ survey volume). The other multi-band combinations 
cover the full range of detection/non-detection combinations that might be expected for T dwarfs.

For comparison, the recent large-scale WISE search made by \citet{kirkpatrick2011} overlaps significantly 
with our search-space. However, they use a slightly bluer $W1-W2\ge$1.5 selection, and where we require 
SNR$>$10 in the $W2$-band they require at least 8 separate detections (SNR$>$3) in the individual $W2$ exposures.

\begin{table*}
\begin{tabular}{|llll|l|l|l|l|l|}
\hline
\multicolumn{4}{|l|}{WISE detection?} & Colours & Selected    & Candidate      & Selected    & Candidate     \\
$W1$ & $W2^a$ & $W3$ & $W4$ &         & sources$^d$ & companions$^{d,e}$ & sources$^f$ & companions$^f$ \\
\hline
n$^b$ & Y$^c$ & n & n & -             & 2418 &  12(3) &  289 & 0 \\
\hline
Y     & Y     & n & n & $W1-W2\ge$2.0 & 5622 &  35(1) & 1330 & 0 \\
\hline
Y     & Y     & Y & n & $W1-W2\ge$2.0 & 1721 &   9(1) &  283 & 0 \\
      &       &   &   & $W2-W3\le$2.5 &      &        &      &   \\
\hline
Y     & Y     & Y & Y & $W1-W2\ge$2.0 & 1018 &   2    & 1642 & 0 \\
      &       &   &   & $W2-W3\le$2.5 &      &        &      &   \\
\hline
n     & Y     & Y & n & $W2-W3\le$2.5 &  174 &   0    &   54 & 0 \\
\hline
n     & Y     & Y & Y & $W2-W3\le$2.5 &  272 &   0    &   54 & 0 \\
\hline
\multicolumn{9}{|l|}{$^a$ $W2$ signal-to-noise always $>$10 (w2snr$>=$10).}\\
\multicolumn{9}{|l|}{$^b$ Non-detections defined as (w$\star$mpro is null or w$\star$sigmpro is null).}\\
\multicolumn{9}{|l|}{$^c$ Detections defined as (w$\star$mpro is not null and w$\star$sigmpro is not null).}\\
\multicolumn{9}{|l|}{$^d$ 2MASS non detections (tmass$\_$key is null).}\\
\multicolumn{9}{|l|}{$^e$ The numbers in brackets are for candidates that passed visual inspection.}\\
\multicolumn{9}{|l|}{$^f$ 2MASS detections with $H-W2\ge$2.5 or $J-W2\ge$3.5.}\\
\end{tabular}
\caption{WISE late T candidate sample. Twelve separate selections were made, six requiring non-detection 
in 2MASS and six requiring 2MASS detection with red 2MASS-WISE colour. The number of candidate T dwarfs, and 
those that became wide companion candidates (see Section 3) are indicated for each search, where various 
combinations of detection and non-detection were explored in the four WISE bands.\label{tab:search}}
\end{table*}

\section{Identifying candidate binary systems}
\label{sec:candbinaries}

To identify candidate binary systems we cross-matched our candidate late-T sample with a list of 
potential primary stars with measured parallax distances, and imposed separation constraints on the potential 
binary pairings as well as absolute magnitude constraints on the candidate T dwarfs (where we assumed a 
common distance for components). The list of potential primary stars was made by combining together 
the latest Hipparcos catalogue \citep{vanleeuwen2007} with the most recent version of the catalogue of nearby stars 
\citep{gliese1991}. Hipparcos provides astrometric measurements (in position, parallax, and annual proper 
motion) with uncertainties in the range 0.7-0.9 milliarcsec (mas) for stars brighter than $V$=9. The catalogue 
is on the ICRS reference system and has proper motions consistent with an inertial system at the level of 
$\pm$0.25mas/yr. The Third Catalogue of Nearby Stars (CNS3) contains information on all known stars within 
25 parsecs based on an extensive literature search during almost four decades.

We required on-sky angular separation of candidate pairs to be $\le$300 arcseconds to reduce contamination from 
random alignments, and physical separation of $\le$10,000AU (where we use the distance of the primary to convert 
angular separation into a physical line-of-sight separation), since the great majority of known wide ultracool 
stellar companions have separations below this limit \citep[e.g.][]{zenghua2010,faherty2010}. In addition 
we use the known distance of each candidate primary to estimate the $M_{W2}$ that the T dwarf candidate would have 
at this distance, and reject any associations where the T dwarf candidate would have $M_{W2}\le$11.5 \citep[i.e. 
targeting late T dwarf companions, e.g. Fig 4 of][where $M_{[4.5]}$ is a good proxy for $M_{W2}$]{sandy2010a}. 
The candidate T dwarf components of the possible binary systems are distributed within our WISE source selections 
as summarised in Table \ref{tab:search}. These T dwarf candidates were visually inspected using the WISE image 
server at the NASA/IPAC Infrared Science Archive, and candidates rejected if the source did not appear point-like 
in any of the bands, formed part of a blended structure, or was clearly an artefact (e.g. part of a diffraction spike).

Five candidate binary systems passed visual inspection. One $W2$-only detected candidate remains an unconfirmed 
interesting candidate without any additional survey data (e.g. UKIDSS, VISTA) to facilitate proper motion 
measurements. The other 4 are listed below:

\begin{itemize}
  \item{WISEP J075003.84+272544.8 is a $W2$-only detected candidate 265 arcseconds from the star Hip 38228, 
a G5IV star at 22pc. This candidate is a known T8.5 dwarf discovered (with WISE) by \citet{kirkpatrick2011}, 
though we subsequently show (Section \ref{sec:propermotion}) that it is not a companion to Hip 38228.}
  \item{WISEP J142320.86+011638.1 (WISEP J1423+0116) is a $W2$-only detected candidate 153 arcseconds 
from the star Hip 70319 (BD+01~2920), a G1V star at 17.2pc. It was not identified by \citet{kirkpatrick2011} 
because it is only detected in 7 separate $W2$ exposures in the WISE Preliminary Data Release. This T dwarf is 
the main subject of this paper.}
  \item{WISEP J145715.85-212207.6 is a $W1+W2+W3$ detected candidate near the system Gl 570ABC 
(Hip 73182 and Hip 73184), a K4V+M1.5V+M3V triple system. This candidate is a known (discovered in 2MASS) 
T8 member of the multiple system \citep{burgasser2000}. The WISE catalogue does not list the source 
as a 2MASS detection because its high proper motion leads to the WISE and 2MASS positions being separated 
by more than 3 arcseconds.}
  \item{WISEP J150457.58+053800.1 is a $W1+W2$ detected candidate 63 arcseconds from Hip 73786 (GJ 576), 
a K8V star at 18.6pc. This candidate is a known (discovered in UKIDSS) T6p companion to this somewhat 
metal poor star \citep{dm2011,scholz2010}.}
\end{itemize}

\begin{figure*}
\includegraphics[height=240pt, angle=0]{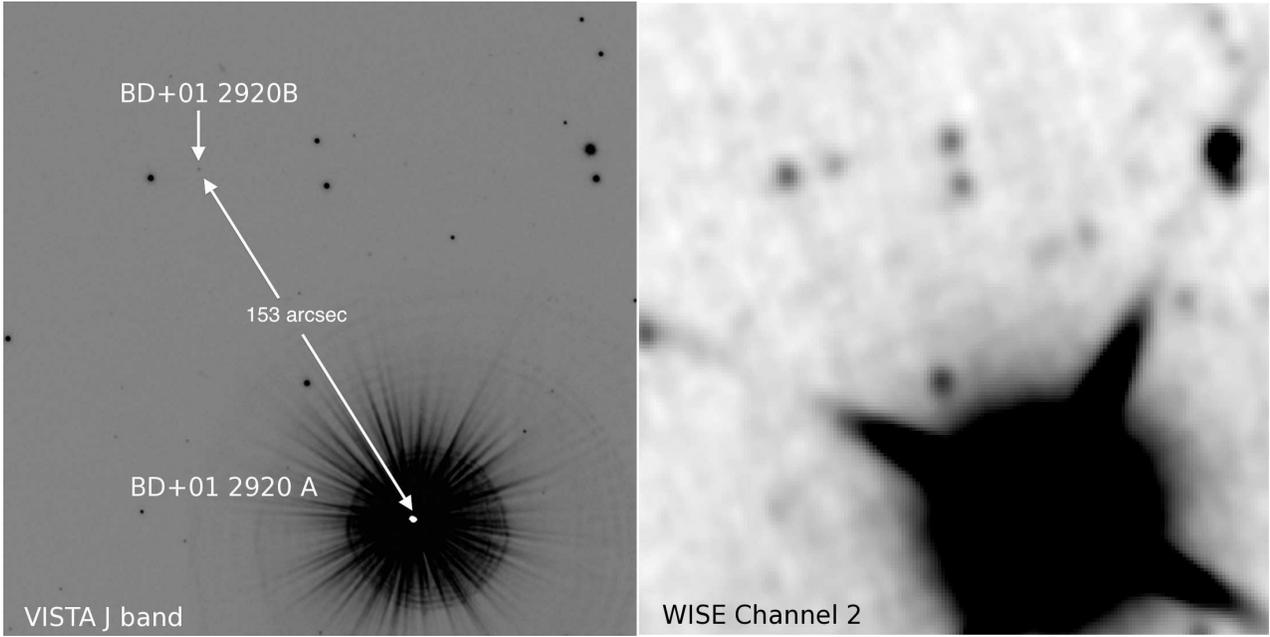}
\caption{$J$-band and $W2$-band images of WISEP J1423+0116.}
\label{fig:finder}
\end{figure*}

Figure \ref{fig:finder} shows $J$- (VISTA) and $W2$-band images for WISEP J1423+0116, and indicates its separation 
from the nearby high proper motion star Hip 70319 (BD+01~2920).

\section{Spectroscopy}
\label{sec:spectra}

\begin{figure*}
\includegraphics[height=500pt, angle=90]{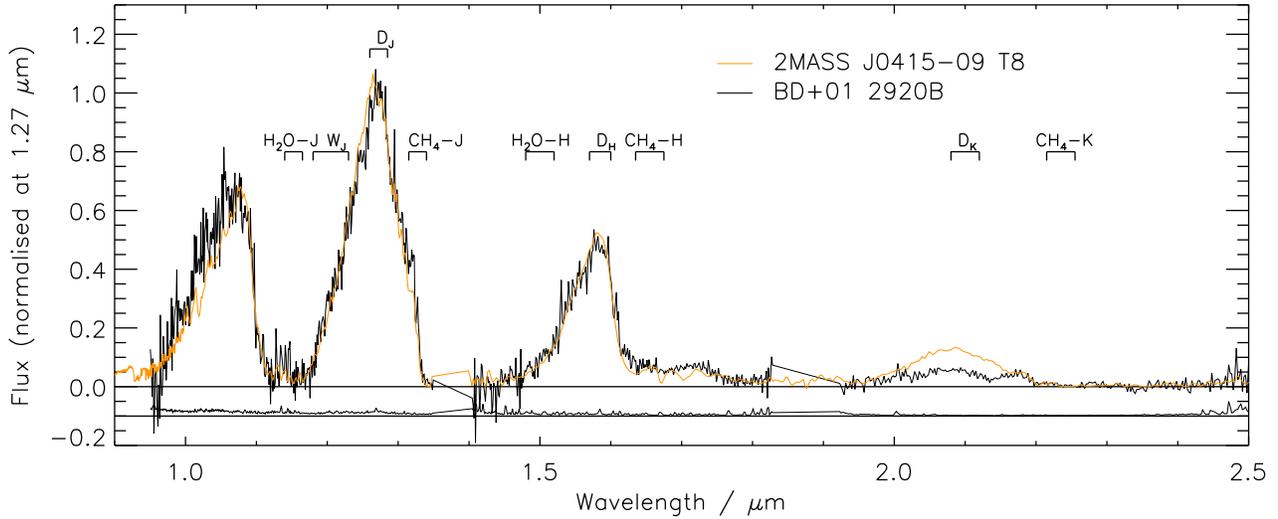}
\caption{GNIRS spectra for WISEP J1423+0116 (BD$+01 2920$B; see Section \ref{sec:binarity}) compared to T8 
spectral type template 2MASS~J04151954-0935066 taken from \citet{burgasser06}. The error spectrum is shown 
offset by $-0.1$.}
\label{fig:jhkspec}
\end{figure*}

Near-infrared spectroscopy of WISEP J1423+0116 (BD$+01 2920$B; see Section \ref{sec:binarity}) was obtained 
using the Gemini Near InfraRed Spectrograph \citep[GNIRS; ][]{elias2006} mounted on the Gemini-North telescope 
on the night of 16$^{th}$ May 2011. The target was observed in cross-dispersed mode capturing the full 
0.8--2.5$\micron$ region with a 1.0$\arcsec$ slit delivering a resolving power of R$\sim 500$. The data were 
reduced using GNIRS routines in the Gemini {\sc IRAF} package \citep{cooke2005}, using the nearby F5V star 
Hip~63976 for telluric correction. The telluric standard spectrum was divided by a black-body spectrum of an 
appropriate $T_{\rm eff}$ after removing hydrogen lines by interpolating the local continuum. The rectified 
standard spectrum was then used to correct for telluric absorption and to provide relative flux calibration. 
The overlap regions between the orders in the $Y-$,$J-$ and $H-$bands agreed well suggesting that the relative 
flux of the orders is well calibrated. The resulting $YJHK$ spectra are shown in Figure~\ref{fig:jhkspec}.

In Figure~\ref{fig:jhkspec} we compare our GNIRS spectrum of the new T8 with that of the T8 spectral template 
2MASS~J04151954-0935066 from \citet{burgasser06}. The close similarity of the spectra over most of the wavelength 
range argues strongly for T8 classification, which is reflected in the values found for the spectral typing flux 
ratios (see Table~\ref{tab:indices}). Although the new T8 closely traces the T8 template over the $0.9 - 1.9 \micron$ 
range, it displays a considerably more depressed $K$-band flux, which is interpreted as due to strong collisionally 
induced absorption by H$_2$ \citep[CIA H$_2$; ][]{saumon94}. Increased CIA H$_2$ is typically attributed to higher 
pressure atmospheres arising from lower-metallicity and/or high-gravity \citep[e.g. ][]{burgasser02, knapp04,liu07}. 
For this reason we assign the type T8p to WISEP J1423+0116, where the 'p' suffix  denotes it is peculiar, alluding 
to the poor match with the template in the $K$ band.

\begin{table}\renewcommand{\arraystretch}{3}\addtolength{\tabcolsep}{-1pt}
\begin{tabular}{c c c c }
  \hline
 {\bf Index} & {\bf Ratio} & {\bf Value} & {\bf Type} \\
\hline
H$_2$O-J & $\frac{\int^{1.165}_{1.14} f(\lambda)d\lambda}{\int^{1.285}_{1.26}f(\lambda)d\lambda }$ & $0.050 \pm 0.003$  & $\geq$T8 \\[+1mm]
CH$_4$-J & $\frac{\int^{1.34}_{1.315} f(\lambda)d\lambda}{\int^{1.285}_{1.26}f(\lambda)d\lambda }$ &  $0.23 \pm 0.01$ & $\geq$T8 \\
$W_J$ & $\frac{\int^{1.23}_{1.18} f(\lambda)d\lambda}{2\int^{1.285}_{1.26}f(\lambda)d\lambda }$   &  $0.32 \pm 0.01$  & T8 \\
H$_2$O-$H$ & $\frac{\int^{1.52}_{1.48} f(\lambda)d\lambda}{\int^{1.60}_{1.56}f(\lambda)d\lambda }$ & $0.21 \pm 0.01$  & T7/8 \\
CH$_4$-$H$ & $\frac{\int^{1.675}_{1.635} f(\lambda)d\lambda}{\int^{1.60}_{1.56}f(\lambda)d\lambda }$ & $0.15 \pm 0.01$ & T7/8 \\
NH$_3$-$H$ &  $\frac{\int^{1.56}_{1.53} f(\lambda)d\lambda}{\int^{1.60}_{1.57}f(\lambda)d\lambda }$ & $0.68 \pm 0.01$ & ... \\
CH$_4$-K &  $\frac{\int^{2.255}_{2.215} f(\lambda)d\lambda}{\int^{2.12}_{2.08}f(\lambda)d\lambda }$ & $0.13 \pm 0.01$ & T6/7 \\
\hline
\end{tabular} 
\caption{The spectral flux ratios for WISEP J1423+0116 (BD$+01 2920$B; see Section \ref{sec:binarity}). The locations of the 
numerators and denominators are indicated on Figure~\ref{fig:jhkspec}.}
\label{tab:indices}
\end{table}

\section{New photometry}
\label{sec:phot}

Database photometry of WISEP J1423+0116 was obtained from the WISE Preliminary Data Release catalogue, 
the WFCAM Science Archive (UKIDSS Large Area Survey) and the VISTA Science Archive (VIKING proprietary 
data access). In addition, observations were taken at the Telescopio Nazionale Galileo (\textsc{tng}) 
and with the Spitzer Space Telescope in its warm phase.

Near-infrared photometry was measured using the Near Infrared Camera Spectrometer \citep[\textsc{nics};][]{baffa2001} 
at the 3.58-m optical/infrared \textsc{tng} located on La Palma, on the night of the 7$^{\mathrm{th}}$ of 
May 2011 for the {\it H} band and the night of the 10$^{\mathrm{th}}$ of June 2011 for the {\it Y} band. 
The data were obtained in large field mode, with a pixel scale of 0.25 arcsec/pixels 
and a field of view of 4.2$\times$4.2 arcmins. The data were processed using the \textsc{nics} science 
pipeline \textsc{snap} provided by the \textsc{tng}. $H$-band observations consisted of a 50 point jitter 
pattern with individual 10s exposures and 6 co-adds per jitter point, accumulating to a total exposure 
time of 50 minutes. In the $Y$-band a 10 point jitter was used for the same exposure time and co-adds, 
resulting in a total exposure time of 10 minutes. We calibrated each image onto the \textsc{mko} system 
using $\sim$30 field stars in the frame.

Warm-Spitzer photometric data were obtained for WISEP J1423+0116 on the 21$^{st}$ August 2011, via the 
Cycle 7 GO program 70058. Individual frame times were 30 s repeated six times, with a 16-position spiral 
dither pattern, for a total integration time of 48 min in each of the $[3.6]$ and $[4.5]$ bands. The 
post-basic-calibrated-data mosaics generated by version 18.18.0 of the Spitzer pipeline were used 
to obtain aperture photometry. The photometry was derived using a 7-arcsec aperture and the aperture 
correction was taken from the IRAC handbook. The error is estimated by the larger of either the variation 
with the sky aperture or the error implied by the uncertainty images.

Tables \ref{tab:mags} and \ref{tab:colours} contain the available photometry and colours, respectively, 
for the T dwarf. We present WISE photometry where the signal-to-noise is positive and note that the 
$W1$ and $W3$ magnitudes are brightness upper limits. The near infrared photometry is on the Mauna-Kea 
Observatory system \citep{sandy2006} except for the TNG $Y$-filter, which is slightly different 
($\lambda_c$=1.02$\micron$, FWHM=0.13$\micron$) to the MKO $Y$ filter ($\lambda_c$=1.02$\micron$, 
FWHM=0.10$\micron$). In the absence of a measured $K$-band magnitude we have determined a synthetic 
$J-K$ colour using our GNIRS spectra and a spectrum of Vega \citep{bohlin2004} both convolved with 
the response functions for the pass-bands \citep[e.g.][]{Hewett06}. This synthetic colour (see Table 
\ref{tab:colours}) combined with the $J$-band magnitude produced our $K$ band estimate. For the mid infrared 
photometry we note that while similar to $W1$ and $W2$, the Spitzer [3.6] and [4.5] bands have some 
significant differences \citep[see Fig 2 of][]{mainzer2011}. There are multiple measurements of $Y$-, 
$J$- and $H$-band photometry, though no evidence of variability is seen (to within the uncertainties) 
in the photometric brightness.

\begin{table*}
\begin{tabular}{|l|l|l|l|l|l|l|l|l|}
\hline
Source & $Y^a$ & $J$ & $H$ & $K$ & $W1$(snr) or $[3.6]$ & $W2$(snr) or $[4.5]$ & $W3$(snr) & $W4$ \\
\hline
WISE & & & & & 17.75(1.5)$^b$ & 14.76$\pm$0.09(11.8) & 12.21(0.8)$^b$ & - \\
UKIDSS LAS & 19.51$\pm$0.14 & 18.76$\pm$0.12 & & \\
VISTA VIKING & 19.69$\pm$0.05 & 18.71$\pm$0.05 & & \\
TNG & 19.75$\pm$0.22 & & 19.14$\pm$0.20 & \\
Synthetic Estimate & & & (18.96$\pm$0.15) & (19.89$\pm$0.33) \\
Spitzer & & & & & 16.77$\pm$0.03 & 14.71$\pm$0.01 & & \\
\hline
\multicolumn{9}{|l|}{$^a$ Photometry is on the MKO system except for the TNG $Y$ filter (see text).}\\
\multicolumn{9}{|l|}{$^b$ 95\% confidence brightness upper limit.}\\
\hline
\end{tabular}
\caption{Photometric magnitudes of WISEP J1423+0116.\label{tab:mags}}
\end{table*}

\begin{table*}
\begin{tabular}{|l|l|l|l|l|l|l|l|l|l|}
\hline
$Y-J$ & $J-H$ & $H-K$ & $J-K$ & $W1-W2$ & $W2-W3$ & $J-W2$ & $H-W2$ & $[3.6]-[4.5]$ & $H-[4.5]$ \\
\hline
0.98$\pm$0.07 & -0.38$\pm$0.23 & -0.93$\pm$0.36$^a$ & -1.27$\pm$0.34$^a$ & $\ge$2.77$^b$ & $\le$2.55$^b$ & 3.95$\pm$0.10$^c$ & 4.38$\pm$0.22 & 2.06$\pm$0.03 & 4.43$\pm$0.20 \\
\hline
\multicolumn{8}{|l|}{$^a$ Synthetic photometry (see text).}\\
\multicolumn{8}{|l|}{$^b$ 95\% confidence limit.}\\
\multicolumn{8}{|l|}{$^c$ Using the higher signal-to-noise VISTA $J$-band magnitude.}\\
\hline
\end{tabular}
\caption{Photometric colours of WISEP J1423+0116.\label{tab:colours}}
\end{table*}

\section{Proper motions}
\label{sec:propermotion}

We measured the proper motion of WISEP J1423+0116 using a VISTA VIKING image of 
April 2010 and two UKIDSS images from May 2008 (with lower signal-to-noise of 
$\sim$6). This avoids using the larger point-spread-functions inherent in the 
WISE images ($\sim$6.5 arcsecs in $W2$). The base-line between the two near infrared 
epochs was 1.89 years. We took the measured x,y coordinates from the standard CASU 
pipeline reductions of all images and using 59 objects within 4 arcminutes of the 
target, transformed the UKIDSS frames onto the standard coordinate system of the 
VIKING frame using a simple linear model. The relative proper motion for all objects 
were found from linear fits to the standard coordinates at the different epochs. 
A correction to an absolute system was estimated from the median difference 
between measured relative proper motions and 6 SDSS objects in the field 
with proper motions in the catalog of \citet{Munn2004}. The derived proper 
motion for WISEP J1423+0116 was corrected for an assumed parallax of
50mas (see Section \ref{sec:binarity}), and final uncertainties are based on the formal 
uncertainties of the measured coordinates combined with an additional 
allowance for the centroiding accuracy in the low signal-to-noise LAS image 
($\sim\pm$0.5 pixels estimated using Monte Carlo techniques) leading to a 
proper motion uncertainty of $\pm$50 mas/yr. The proper motion of WISEP 
J1423+0116 is $\mu_{\alpha \cos{\delta}}=261\pm 56$ mas~yr$^{-1}$, 
$\mu_{\delta}=-444\pm 52$ mas~yr$^{-1}$, which is within 0.7$\sigma$ of the 
Hipparcos proper motion vector of Hip 70319 (BD+01 2920; 
$\mu_{\alpha \cos{\delta}}=223.8\pm 0.4$ mas~yr$^{-1}$, $\mu_{\delta}=-477.4\pm 0.4$ mas~yr$^{-1}$). 
These objects are thus a common proper motion pair.

We also measured the proper motion of WISEP J075003.84+272544.8 using two UKIDSS LAS 
J-band epochs with a baseline of 2 years. We applied a second order polynomial 
transformation between the two epoch images to correct for any non-uniformity in 
the focal plane. Seventeen reference stars ($J<$18.1) were used, distributed around 
the target (with at least 3 per quadrant) with separations within 2 arcminutes. A 
correction was applied to an absolute system using apparent proper motions of nearby 
galaxies. The uncertainties were calculated using the standard deviations in the RA/Dec 
residuals of sources deemed to have insignificant motion ($<$ 45 milli-arcseconds) between 
epochs. The proper motion of WISEP J075003.84+272544.8 is 
$\mu_{\alpha \cos{\delta}}=-732\pm 17$mas~yr$^{-1}$, $\mu_{\delta}=-194\pm 17$mas~yr$^{-1}$. 
By comparison, Kirkpatrick et al. (2011) used astrometric fits to multiple WISE 
observations to derive a proper motion ($\mu_{\alpha \cos{\delta}}=-869\pm 424$mas~yr$^{-1}$, 
$\mu_{\delta}=-1107\pm 438$mas~yr$^{-1}$) with much larger uncertainties. Their value 
of $\mu_{\alpha \cos{\delta}}$ is consistent with our new measurement, however their 
value of $\mu_{\delta}$ is too large at the level of $\sim$2$\sigma$. Although there 
also happens to be a nearby Hipparcos star (Hip 38228), it has a low proper motion 
($\mu_{\alpha \cos{\delta}}=-8$mas~yr$^{-1}$, $\mu_{\delta}=-10$mas~yr$^{-1}$) and the 
T dwarf is not a common proper motion companion since its motion differs at a level 
$>$28$\sigma$.

\section{Confirming binarity}
\label{sec:binarity}

To estimate the probability that WISEP J1423+0116 and BD+01 2920 may be 
a line-of-sight association as opposed to a genuine binary, we have performed 
a statistical analysis to estimate the expected number of chance alignments in 
our search. We used the \citet{ben2010b} luminosity function constraints 
to estimate the number of T6-9 dwarfs expected in the WISE first data release. 
In a sample with $W2<$15 (akin to our $W2$ signal-to-noise requirement) we 
expect to detect T7$\pm$1 and T9 dwarfs out to distances ($D_{max}$) of $\sim$25 
and $\sim$15pc respectively, in the 57\% sky coverage of the WISE first release. 
We adjusted the Burningham luminosity function to add back in the correction made 
for unresolved binarity (3-45\%), since this removed T dwarfs from their magnitude 
limited samples, and estimate an expected 28-251 T7$\pm$1 dwarfs and 26-111 T9 
dwarfs in the WISE selection using this luminosity function. We then summed the 
volume in which line-of-sight associated stars may be found using a set of cones 
(one per T dwarf) each with its apex at the observer and a T dwarf in the centre 
of it's base (using a base radius of 10,000AU to match our search criteria). Since 
the number of T dwarfs is proportional to D$^3$ and the volume of a cone is proportional 
to D (where D is distance), the average cone volume will be $\frac{3}{4}$ of the 
maximum cone volume $\frac{1}{3}AD_{max}$ (where $A$ is the base area of a cone 
$\pi\times 10,000AU^2$). The total cone volume for T6--9 dwarfs was thus estimated 
to be 2.0--14.6pc$^3$.

The luminosity function of \citet{reid2007} for the 8pc and 20pc samples leads 
to a stellar density of 0.062--0.076 stars pc$^{-3}$, and we thus expect 0.12--1.11 
light-of-sight associations between stars and T dwarfs in our WISE selection. 
Amongst our five candidates we find that one of them (WISEP J075003.84+272544.8) 
is in fact a line-of-sight association with a lack of common proper motion. This 
is consistent with our estimates above. An additional candidate was identified 
without proper motion, though the above statistic does not provide any further 
indications on the likelihood that this candidate may be genuine.

We must also assess these common proper motion systems for the chance that common 
proper motions are aligned by random chance. Using the proper motion and direction 
of WISEP J1423+0116 we estimated this probability using a Hipparcos sample, 
downloading the proper motions of Hipparcos stars within 45 degrees of the 
WISEP J1423+0116/BD+01 2920 pair, and with distances from 10--40pc (the photometric 
distance range of a T8$\pm$1 dwarf with $J\simeq$18.7 allowing for possible 
unresolved binarity). We counted the fraction of stars with proper motion 
within 55mas~yr$^{-1}$ (1$\sigma$) of the T dwarf motion, and thus estimate a chance 
of 1.3\% that this high proper motion pair could be common proper motion by random 
chance. We therefore expect no more than 0.0015--0.014 false positive common proper 
motion systems in the search that we have made, and conclude that all three of the 
common proper motion systems that we identified are genuine binaries. This includes 
the two previously reported systems and the association between WISEP J1423+0116 
and BD+01 2920, which becomes the binary system BD+01 2920AB.

\section{Properties of BD+01~2920A}
\label{sec:primary}

A search of the literature reveals multiple studies of the primary star BD+01~2920A. 
It is a nearby high proper motion G1 dwarf (0.9M$_{\odot}$) with thin disk kinematics. 
There is no evidence of any debris disk or low-mass companions (including giant planets), 
and it has low activity. BD+01~2920A is a mildly metal poor star, with a metallicity in 
the metal poor tail of the disk distribution rather than in the halo regime. With one 
exception previous estimates of [Fe/H] are in the range -0.38$\pm$0.06 
\citep[only][gives a slightly higher metallicity of -0.20 dex]{lebreton1999}. 
The range of age constraints covers 2.3--14.4 Gyr. This differs slightly from the range 
quoted by \citet{lawler2009} who give a lower limit of 3.5 Gyr. The difference is due 
to the estimate of 2.27 Gyr from \citet{donascimento2010}, and we here adopt an inclusive 
age range. The properties of BD+01~2920A are summarised in Table \ref{tab:primary} and 
references therein.

\begin{table*}
\begin{tabular}{|l|l|}
\hline
BD+01 2920A (Hip 70319) & \\
\hline
R.A. (J2000) &  14 23 15.285 \\
Dec (J2000)  & +01 14 29.65 \\
PM$_{\alpha \cos{Dec}}$ & $223.8\pm0.4$ mas/yr \\
PM$_{Dec}$ & $-477.4\pm0.4$ mas/yr \\
Spectral type/class & G1V \\
$\pi$ & 58.2$\pm$0.5 mas  $^{(1)}$ \\
Distance & 17.2$\pm$0.2 pc \\
$m-M$ & 1.18$\pm$0.03 \\
V$_r$ & 19.6 $\pm$ 0.3 km/s$^{(2-4)}$ \\
Space motion & UVW = 22, 15, 39 $^{(5-8)}$ \\
Population & Thin disk $^{(9,10)}$ \\
$T_{\rm eff}$ & 5750 $\pm$ 100 K $^{(3,7,8,11-21)}$ \\
$\log{g}$ & 4.45 $\pm$ 0.05 dex $^{(3,7,12,16-18,20-24)}$ \\
Mass & 0.87 $\pm$ 0.07 M$_{\odot}$ $^{(3,13,23)}$ \\
$[$Fe/H$]$ & -0.38 $\pm$ 0.06 dex $^{(3,6-8,10,12,14-18,20-29)}$\\
Age & 2.3 -- 14.4 Gyr $^{(3,5,8,10,12,13,15,23,26,30-32)}$ \\
v$\sin{i}$ & 1--2 km/s $^{(3,12,13)}$ \\
Activity & Low activity star $^{(33)}$ \\
Debris disk & None $^{(15)}$ \\
Close in companions & No $\ge$70-75M$_{Jup}$ at 20-250AU $^{(34)}$ \\
                    & No giant planets \\
                    & ($>$100m/s) at $<$ 5AU $^{(24,35-39)}$ \\
\hline
\multicolumn{2}{|l|}{$^1$ \citet{vanleeuwen2007}, $^2$ \citet{gontcharov2006}, $^3$ \citet{valenti2005}}\\
\multicolumn{2}{|l|}{$^4$ \citet{latham2002}, $^5$ \citet{holmberg2009}, $^6$ \citet{ramirez2007}}\\
\multicolumn{2}{|l|}{$^7$ \citet{mishenina2004}, $^8$ \citet{nordstroem2004}, $^9$ \citet{borkova2004}}\\
\multicolumn{2}{|l|}{$^{10}$ \citet{ibukiyama2002}, $^{11}$ \citet{casagrande2010}, $^{12}$ \citet{takeda2010}}\\
\multicolumn{2}{|l|}{$^{13}$ \citet{donascimento2010}, $^{14}$ \citet{holmberg2007}, $^{15}$ \citet{lawler2009}}\\
\multicolumn{2}{|l|}{$^{16}$ \citet{luck2006}, $^{17}$ \citet{shi2004}, $^{18}$ \citet{mashonkina2007}}\\
\multicolumn{2}{|l|}{$^{19}$ \citet{kovtyukh2003}, $^{20}$ \citet{giridhar2002}, $^{21}$ \citet{cayreldestrobel2001}}\\
\multicolumn{2}{|l|}{$^{22}$ \citet{wu2011}, $^{23}$ \citet{takeda2007b}, $^{24}$ \citet{heiter2003}}\\
\multicolumn{2}{|l|}{$^{25}$ \citet{mashonkina2001}, $^{26}$ \citet{rochapinto1998}, $^{27}$ \citet{karatas2005}}\\
\multicolumn{2}{|l|}{$^{28}$ \citet{borkova2005}, $^{29}$ \citet{haywood2001}, $^{30}$ \citet{wright2004}}\\
\multicolumn{2}{|l|}{$^{31}$ \citet{barry1988}, $^{32}$ \citet{takeda2007a}, $^{33}$ \citet{hall2007} $^{34}$ \citet{carson2009}}\\
\multicolumn{2}{|l|}{$^{35}$ \citet{halbwachs2003}, $^{36}$ \citet{halbwachs2000}, $^{37}$ \citet{cumming1999}}\\
\multicolumn{2}{|l|}{$^{38}$ \citet{endl2002}, $^{39}$ \citet{nidever2002}}\\
\hline
\end{tabular}
\caption{Properties of BD+01 2920A (Hip 70319).\label{tab:primary}}
\end{table*}

\section{Properties of BD+01~2920B}
\label{sec:properties}

\subsection{Bolometric flux}
\label{subsec:boloflux}

We estimate the bolometric flux ($F_{bol}$) of the new T8p companion BD+01~2920B following a similar method to that 
outlined in \citet{ben2009}, by combining our $YJHK$ spectrum (flux calibrated by our follow-up photometry) with model 
spectra (to allow us to estimate the flux contributions from regions outside our near-infrared spectral coverage). 
We have scaled the $\lambda < 1.05 \micron$ region of the models to match the flux level in our $YJHK$ spectrum, 
and we have used the Spitzer 3.6$\micron$ and 4.5$\micron$ photometry to scale the $\lambda = 2.5-3.95\micron$ 
and $\lambda > 3.95\micron$ regions respectively (the transmission profiles of the Spitzer filters cross at 
3.95 $\micron$ at a transmission level of $\sim$1\%). To avoid biasing our derived flux estimate with our choice 
of model, we have produced estimates using Solar and [M/H]$ = -0.3$ metallicity BT Settl models \citep{PTRSA_water} 
that bracket the likely range of gravities and $T_{\rm eff}$ for our target ($\log g = 4.50-5.5$; $T_{\rm eff} = 
500-900$K). We take the scatter in the estimates resulting from different model choices as a reflection of the 
systematic uncertainty introduced by the atmospheric models. We have used a Monte Carlo method to determine the 
uncertainty in each estimate due to the noise in the photometry used for scaling the models and the noise in our 
GNIRS spectrum. Our final estimate of $F_{bol}$ is the median of our estimates using different model extensions, 
whilst the uncertainty is the sum in quadrature of the systematic uncertainty and the mean random uncertainty. 
This results in a determination of $F_{bol} = 1.61 \pm 0.17 \times 10^{-16}$~Wm$^{-2}$.

\subsection{Luminosity, mass, radius, and effective temperature}
\label{subsec:lmrt}

The luminosity of BD+01~2920B was derived from the bolometric flux and the distance. The on-sky separation of the 
BD+01~2920AB components leads to a tangential separation of 0.01 pc. This is negligible compared with the uncertainty 
in the parallax distance of the primary ($\pm$0.2 pc) and we can therefore assume that the T dwarf is at the same 
distance as BD+01~2920A (17.2$\pm$0.2 pc). Taking the uncertainties associated with the bolometric flux and distance 
into account leads to the determination of $L_{bol}=5.69\pm 0.60\times 10^{-20}$W, or $\log{L/L_{\odot}}=-5.83\pm 0.05$. 
To determine the $T_{\rm eff}$ of BD+01~2920B we estimate its radius using the COND evolutionary models \citep{baraffe2003}. 
These models reproduce the main trends of observed methane dwarfs in near-infrared color-magnitude diagrams, though are 
only available for solar metallicity.

We used linear interpolation between the model isochrones to estimate a range of possible mass, radii and surface 
gravity for BD+01~2920B consistent with an age range of $\sim$2--10 Gyr. Accounting for the uncertainties in the 
measured luminosity we obtained a mass range of 0.019--0.047 M$_{\odot}$ ($20-50$ M$_{Jup}$), a radius range 0.080$-$0.099 
R$_{\odot}$ (0.80--0.99 R$_{Jup}$), and a surface gravity range of $\log{g}=4.68-5.30$. The corresponding temperature 
(from luminosity and radius) is $T_{\rm eff}=680\pm 55$K. A summary of the properties of BD+01~2920B is given in 
Table \ref{tab:secondary}.

\begin{table}
\begin{tabular}{|l|l|}
\hline
\multicolumn{2}{|l|}{BD+01 2920B (WISEP 1423+0116)}\\
\hline
R.A. (J2000) & 14 23 20.86 \\
Dec (J2000)  & +01 16 38.1 \\
PM$_{\alpha \cos{Dec}}$ & $261\pm56$ mas~yr$^{-1}$ \\
PM$_{Dec}$ & $-444\pm52$ mas~yr$^{-1}$ \\
Spectral type & T8p \\
Separation & 153 arcsecs \\
           & 2630 AU $^a$ \\
$F_{bol}$ & (1.61$\pm$0.17)$\times$10$^{-16}$ W$m^{-2}$ $^b$ \\
$m-M$ & 1.18$\pm$0.02$^a$ \\
$M_Y$ & 18.51$\pm$0.04 $^a$ \\
$M_J$ & 17.53$\pm$0.05 $^a$ \\
$M_H$ & 17.96$\pm$0.20 $^a$ \\
$M_K$ & 18.71$\pm$0.33 $^{a,c}$ \\
$M_{3.6}$ & 15.59$\pm$0.04 $^a$ \\
$M_{4.5}$ & 13.53$\pm$0.02 $^a$ \\
$\log{L/L_{\odot}}$ & $-$5.83 $\pm$ 0.05 $^a$ \\
$[Fe/H]$ & $-$0.38 $\pm$ 0.06 dex $^d$ \\
Mass & 0.019--0.047 M$_{\odot}$ $^e$ \\
     & 20-50 M$_{Jup}$ $^e$ \\
Radius & 0.080--0.099 R$_{\odot}$ $^e$ \\
       & 0.80--0.99 R$_{Jup}$ $^e$ \\
$\log{g}$ & 4.68--5.30 dex $^{e}$ \\
$T_{\rm eff}$ & 680$\pm$55 K $^{f}$ \\
\hline
\multicolumn{2}{|l|}{$^a$ Inferring a distance of 17.2$\pm$0.2pc from BD+01 2920A.}\\
\multicolumn{2}{|l|}{$^b$ Integrating measured flux from $1.0–-2.4 \micron$ and adding a}\\
\multicolumn{2}{|l|}{theoretical correction at other wavelengths (see text).}\\
\multicolumn{2}{|l|}{$^c$ Synthetic photometry used (see text).}\\
\multicolumn{2}{|l|}{$^d$ Inferred from BD+01 2920A.}\\
\multicolumn{2}{|l|}{$^e$ Constraints derived from structure models as a function of}\\
\multicolumn{2}{|l|}{luminosity for ages 2–10 Gyr.}\\
\multicolumn{2}{|l|}{$^f$ Derived from the luminosity and radius constraints.}\\
\hline
\end{tabular}
\caption{Properties of BD+01 2920B (WISEP 1423+0116).\label{tab:secondary}}
\end{table}

Observations of transiting very low-mass stars and brown dwarfs with mass$>$20 M$_{Jup}$ 
\citep{pont2005a,pont2005b,deleuil2008,bouchy2011a,anderson2011,johnson2011} are all 
consistent with the COND mass-radius model data \citep[see fig 10 of][]{bouchy2011b}. These systems are solar 
metallicity to within the uncertainties, though some of these uncertainties are significant. The effects of 
metallicity on sub-stellar radii are a little unclear. \citet{burrows2011} present evolutionary models with 
a spread in radius (at a given mass and age) of $\sim$10--25\%, with higher-metallicity (higher-cloud-thickness) 
atmospheres giving larger radii. However, a comparison between KOI-423b and CoRoT-3b suggests the converse trend. 
KOI-423b orbits a metal poor star ([Fe/H]=$-0.29\pm$0.10) and is a relatively large (1.22$^{+0.12}_{-0.10}R_{Jup}$) 
brown dwarf (18 M$_{Jup}$), whereas CoRoT-3 is a solar metallicity star ([Fe/H]=-0.02$\pm$0.06) hosting 
a smaller (1.01$\pm$0.07R$_{Jup}$) brown dwarf (22M$_{Jup}$). Given this uncertainty in the radius-metallicity 
trend, we do not attempt to make a metallicity correction to our radius estimate. We note however that our COND 
radius constraint already includes an uncertainty at the level of 25\%, comparable with the size of the theoretical 
trends suggested by the \citet{burrows2011} models for a metallicity difference of 1.0 dex.

\begin{figure}
\includegraphics[height=400pt, angle=0]{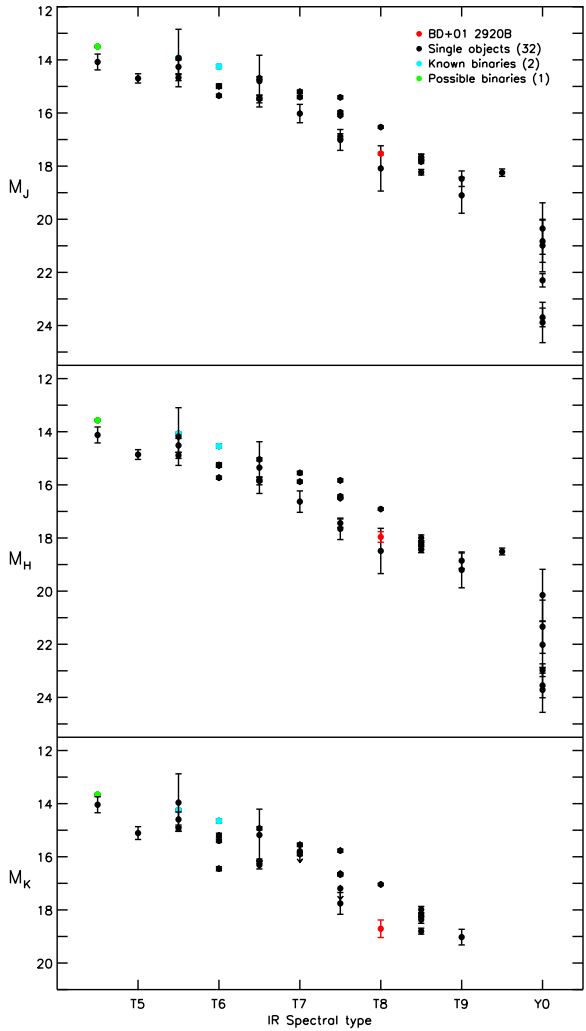}
\caption{Absolute magnitude ($M_{J,H,K}$) spectral type plots showing BD+01~2920B (red symbol) amongst the existing 
population of mid-late T dwarfs and Y dwarfs. All the T dwarfs with spectral types $\le$T8.5 \citep{ben2008} have 
distances from parallax. The T9.5 and Y dwarfs all have spectroscopic distances \citep[except for WISEP J1541-2250 
which has a parallax in][]{kirkpatrick2011}, estimated via $T_{\rm eff}$ and $\log{g}$ constraints from model fits 
to near-infrared spectroscopy.}
\label{fig:absmagplots}
\end{figure}

As an additional caveat we note that our $T_{\rm eff}$ determination relies on an assumption that the object is single, 
and not an unresolved binary. Unresolved binarity would lead to lower $T_{\rm eff}$ for each unresolved component. 
If BD+01~2920B is an equal luminosity unresolved binary the $T_{\rm eff}$ of each component would be a factor $\sim$0.8 
less ($\frac{1}{2}^{\frac{1}{4}}$ with similar radii for the components). For unequal luminosity components the 
brighter component $T_{\rm eff}$ would be closer to 680K with the fainter one $<$540K. Observations suggest 
\citep[e.g.][]{burgasser2005} that the binary fraction of brown dwarfs (resolved at $\sim$0.1 arcsec resolution) 
in widely separated stellar -- brown dwarf multiples is notably higher (45$\pm$15 per cent) than that of field brown 
dwarfs (18$\pm$7 per cent), and unresolved binaries can also have separation closer than 0.1 arcseconds 
\citep[see][and references therein]{ben2009}. In Figure \ref{fig:absmagplots} we show BD+01~2920B in absolute 
magnitude ($M_{J,H,K}$) spectral type diagrams, along with the known population of late T and Y dwarfs (see caption). 
The $K$-band suppression is evident in the $M_K$ plot, though we also note that there is no clear indication that 
the object is an unresolved binary (e.g. with components of near equal brightness) in the $M_J$ and $M_H$ plots. 
We cannot with high confidence however, rule out the possibility that BD+01~2920B is an unresolved binary.

\section{A metal poor benchmark T8 dwarf}
\label{sec:discussion}

\begin{figure}
\includegraphics[height=240pt, angle=90]{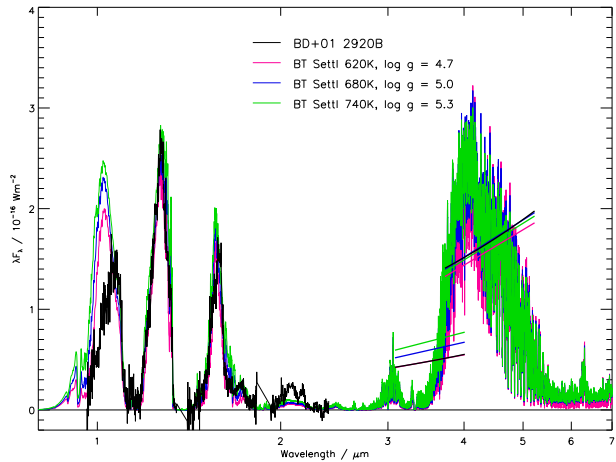}
\caption{The $YJHK$ spectrum of BD$+01~2920$B and mean fluxes inferred from the Spitzer photometry compared
to model spectra that straddle the properties estimated in Section~\ref{subsec:lmrt}. The straight coloured
lines indicate the mean fluxes of model spectra in the IRAC channel 1 and 2  photometric bands and are plotted 
to allow comparison with the mean flux from the target (straight black lines) through the same filters.}
\label{fig:speccomp}
\end{figure}

We now assess some implications of this benchmark system under the assumption that it is a single object, 
and through comparison of its spectrum and colours to theoretical predictions and other ultra-cool objects. 
In Figure~\ref{fig:speccomp} we compare our flux calibrated GNIRS spectrum of BD$+01~2920$B and warm-{\it Spitzer} 
photometry to mildly metal poor BT Settl models \citep{btsettlCS16} for our derived properties, each scaled to 
their corresponding radii and the known distance to the primary star. The BT Settl atmospheric model grid 
spans the cool stellar to substellar temperature regime using the BT2 water line-list \citep{barber2006} 
and the reference Solar abundances of \citet{asplund2009}. In these models dust formation, cloud behaviour 
and vertical mixing are parameterised with reference to the 2D radiation hydrodynamic simulations of 
\citet{bdco5bold}. The $K-$band suppression that is predicted by the models for high-gravity and metal 
poorer brown dwarfs is seen in our GNIRS spectrum of this benchmark T dwarf, although the models predict 
this effect to be stronger than is seen in this case. Similarly poor matches to the observed $K-$band 
spectroscopy have been seen in other benchmark systems \citep[e.g.][]{ben09,ben11a}, so it is reasonable 
to interpret this as a deficiency in the model atmospheres, although its origin is not yet understood. 
It is noteworthy that the $Y-$band spectrum which has also been proposed as diagnostic of metallicity 
variations \citep{bbk2006} is also poorly matched by the models. The model atmospheres provide good 
matches to the flux in the $J-$ and $H-$bands, despite known deficiencies in the methane line lists in 
these regions. The model predictions for the [3.6] and [4.5] fluxes are consistent with those that we 
have observed, and it interesting to note that in the [3.6] band, the 620K, $\log g = 4.7$, model fits 
the data best, whilst in the [4.5] band the two warmer models provide the best match.

\begin{figure}
\includegraphics[height=230pt, angle=90]{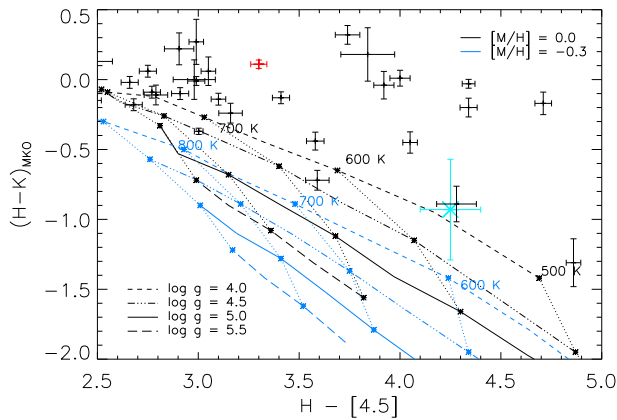}
\caption{Near- to mid-infrared colours of cool T~dwarfs compared to those of 
the BT Settl model colours. BD+01~2920B is indicated with a cyan star symbol, 
whilst the young benchmark Ross 458C is indicated in red.}
\label{fig:hs2plot}
\end{figure}

\begin{figure}
\includegraphics[height=230pt, angle=90]{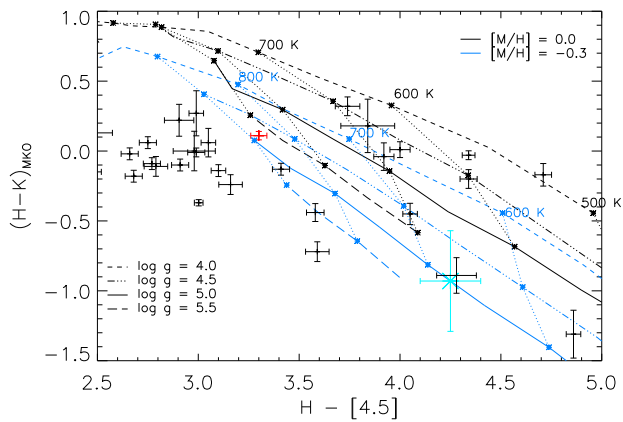}
\caption{Near- to mid-infrared colours of cool T dwarfs compared to those 
of the BT Settl model colours anchored to our estimated properties for 
BD+01 2920B. BD+01 2920B is indicated with a cyan star symbol, whilst the 
young benchmark Ross 458C is indicated in red.}
\label{fig:hs2shift}
\end{figure}

In Figure~\ref{fig:hs2plot} we have plotted synthetic colours in $H-K$ and $H-{\rm [4.5]}$ for the BT Settl 
models along with colours of late T dwarfs from \citet{sandy10} with MKO and IRAC photometry, and 
BD$+01~2920$B.  It can be seen that BD+01~2920B lies in a similar region as other suspected mildly metal-poor 
T7.5/8 dwarfs 2MASS0939 and SDSS1416B, which have Teff 500-700K \citep[e.g.][]{ben2010a,sandy2010a}. 
The effect of the poor match between the models and the data in the $K-$band is highlighted by the 
non-coincidence of the models and the observations in this colour space for the T dwarfs with the 
reddest $H-[4.5]$ colours. However, the models correctly predict the colours for the young benchmark 
Ross 458C. To provide an alternative comparison between the models and the data we have shifted the 
model colours such that they match the observed colours for BD$+01~2920$B for the parameters derived 
in Section~\ref{sec:properties}. Figure~\ref{fig:hs2shift} compares these adjusted model colours to 
the same T dwarfs shown in Figure~\ref{fig:hs2plot}. This plot broadly reproduces the result of \citet{sandy10}, 
where it was noted that the majority of the coolest T dwarfs appear to have low-gravity and/or 
high-metallicity, suggesting that the sample is dominated by young low-mass brown dwarfs (ages 
$\sim 1$Gyr). However, we note that the sources with bluer $H-[4.5]$ lie well below the adjusted 
model tracks in Figure~\ref{fig:hs2shift}, including the young benchmark Ross~458C \citep[for which 
$\log g = 4.0 - 4.7$;][]{ben2011a}, which highlights that a simple offset correction to the models is 
not sufficient to allow the properties of the T dwarf population to be reliably assessed through 
reference to these model colours. The similar temperature of these two benchmarks ($T_{\rm eff} 
\sim 700$K), but wide separation in Figures~\ref{fig:hs2plot} and~\ref{fig:hs2shift} highlights 
the important roles that gravity and metallicity play in determining the $H-K / H - [4.5]$ colours 
for cool T dwarfs.

\section{Conclusions and future work}
\label{sec:conc}

Our cross-match between the WISE first data release and the Hipparcos and Gliese catalogues has resulted in the 
discovery of a new late T binary companion (BD$+01 2920$B) and the re-discovery of two previously known systems. 
WISE (in combination with UKIDSS and VISTA) is thus effectively probing an increased volume of very low 
temperature parameter-space for benchmark companions. There are also significant advantages that the primary 
star BD$+01 2920$A is a nearby G dwarf rather than one of the more numerous M dwarfs in the solar neighbourhood, 
and its lower metallicity provides a crucial test for the effects of reduced metal content on models atmospheres. 
The metallicities and abundances of bright Sun-like stars can be studied with much more confidence and in much 
greater detail than those of M dwarfs, and late T dwarfs in such systems offer the opportunity not only to test 
cool brown dwarf atmosphere physics, but also to potentially study brown dwarf abundances.

In the near future, high resolution imaging (e.g. adaptive optics) observations of BD$+01 2920$B will be 
important to constrain multiplicity on a $\sim$0.1 arcsec ($\sim$1.7AU) separation scale. A close binary 
would be able to provide future dynamical masses (since the orbital period would be just a few years). 
Higher resolution spectroscopy would also be able to assess multiplicity at closer separation, and confirmation 
of a single object nature would validate the approach taken here to determination the physical properties 
of this benchmark object. The existing constraints on the physical properties of BD$+01 2920$B will be 
improved as we develop a better understanding of how brown dwarf radii depend on composition. This will be 
aided by an increasing number of transiting brown dwarfs from Kepler and other transit surveys 
\citep[e.g.][]{borucki2011,me2005}, and improved metallicity measurements for this sample.

We can expect additional late T benchmarks in the future all-sky WISE data release, and a more encompassing 
search of WISE, UKIDSS and VISTA (including at wider angular separations) should yield an expanded population of 
benchmarks across the full T dwarf $T_{\rm eff}$ range. As greater survey volumes are searched for benchmark 
brown dwarfs we can also expect to identify systems with more accurately known ages. Evolved subgiants for 
example are less numerous that their main sequence counterparts, but evolutionary model comparisons can provide 
more accurate age constraints \citep[e.g. $\pm$10\%;][]{me2006}. And as the range of well measured benchmarks 
expands into greater parameter-space we will have the opportunity to comprehensively test the atmosphere models 
by directly mapping the benchmark population's spectral variations/trends onto a grid of tightly constrained 
physical properties.

\section*{Acknowledgments}

This publication makes use of data products from the Wide-field Infrared Survey Explorer, which is a 
joint project of the University of California, Los Angeles, and the Jet Propulsion Laboratory/California 
Institute of Technology, funded by the National Aeronautics and Space Administration.
The UKIDSS project is defined in Lawrence et al. (2007). UKIDSS uses the UKIRT WFCAM (Casali et al. 2007) 
and a photometric system described in Hewett et al. (2006). The pipeline processing and science archive 
are described in Irwin et al. (2004) and Hambly et al. (2008).
Based on observations obtained at the Gemini Observatory, which is operated by the
Association of Universities for Research in Astronomy, Inc., under a cooperative agreement
with the NSF on behalf of the Gemini partnership: the National Science Foundation (United
States), the Science and Technology Facilities Council (United Kingdom), the
National Research Council (Canada), CONICYT (Chile), the Australian Research Council
(Australia), Minist\'{e}rio da Ci\^{e}ncia, Tecnologia e Inova\c{c}\~{a}o (Brazil) 
and Ministerio de Ciencia, Tecnolog\'{i}a e Innovaci\'{o}n Productiva  (Argentina).
This work made use of data obtained on Gemini projects GN-2011A-Q-73.
Based in part on observations made for the VIKING survey, using VISTA at the ESO Paranal Observatory under 
programme ID 179.A-2004. The VISTA Data Flow System pipeline processing and science archive are described 
in Irwin et al (2004) and Hambly et al (2008).
Based on observations made with the Italian Telescopio Nazionale Galileo (A22TAC96) operated on the 
island of La Palma by the Fundaci\'{o}n Galileo Galilei of the INAF (Istituto Nazionale di Astrofisica) 
at the Spanish Observatorio del Roque de los Muchachos of the Instituto de Astrofisica de Canarias.
DP, NL, PL, MCG and ZZ have received support from RoPACS during this research and JG is supported by 
RoPACS, a Marie Curie Initial Training Network funded by the European Commission’s Seventh Framework Programme.
NL is funded by the national program AYA2010-19136 funded by the Spanish ministry of science and innovation.
SKL is supported by the Gemini Observatory, which is operated by AURA, on behalf of the international 
Gemini partnership of Argentina, Australia, Brazil, Canada, Chile, the United Kingdom, and the United 
States of America.
ADJ is supported by a Fondecyt Postdoctorado under project number 3100098.
JSJ is supported by a Fondecyt Postdoctorado under project number 3110004 and partial support from 
Centro de Astrof\'{i}sica FONDAP 15010003, the GEMINI-CONICYT FUND and from the Comit\'{e} Mixto 
ESO-GOBIERNO DE CHILE.
This research has made use of the SIMBAD database, operated at CDS, Strasbourg, France.

\bibliographystyle{mn2e}
\bibliography{refs}

\begin{thebibliography}{114}
\expandafter\ifx\csname natexlab\endcsname\relax\def\natexlab#1{#1}\fi

\bibitem[{{Allard} {et~al}\mbox{.}(1997){Allard}, {Hauschildt}, {Alexander}, \&
  {Starrfield}}]{allard1997}
{Allard} F., {Hauschildt} P.~H., {Alexander} D.~R., {Starrfield} S., 1997,
  \araa, 35, 137

\bibitem[{{Allard} {et~al}\mbox{.}(2001){Allard}, {Hauschildt}, {Alexander},
  {Tamanai}, \& {Schweitzer}}]{allard2001}
{Allard} F., {Hauschildt} P.~H., {Alexander} D.~R., {Tamanai} A., {Schweitzer}
  A., 2001, \apj, 556, 357

\bibitem[{{Allard} {et~al}\mbox{.}(2010){Allard}, {Homeier}, \&
  {Freytag}}]{btsettlCS16}
{Allard} F., {Homeier} D., {Freytag} B., 2010, ArXiv e-prints

\bibitem[{{Allard} {et~al}\mbox{.}(2011){Allard}, {Homeier}, \&
  {Freytag}}]{PTRSA_water}
{Allard} F., {Homeier} D., {Freytag} B., 2011, PTRSA, accepted

\bibitem[{{Anderson} {et~al}\mbox{.}(2011){Anderson}, {Collier Cameron},
  {Hellier}, {Lendl}, {Maxted}, {Pollacco}, {Queloz}, {Smalley}, {Smith},
  {Todd}, {Triaud}, {West}, {Barros}, {Enoch}, {Gillon}, {Lister}, {Pepe},
  {S{\'e}gransan}, {Street}, \& {Udry}}]{anderson2011}
{Anderson} D.~R. {et~al.}, 2011, \apjl, 726, L19

\bibitem[{{Asplund} {et~al}\mbox{.}(2009){Asplund}, {Grevesse}, {Sauval}, \&
  {Scott}}]{asplund2009}
{Asplund} M., {Grevesse} N., {Sauval} A.~J., {Scott} P., 2009, \araa, 47, 481

\bibitem[{{Baffa} {et~al}\mbox{.}(2001){Baffa}, {Comoretto}, {Gennari}, {Lisi},
  {Oliva}, {Biliotti}, {Checcucci}, {Gavrioussev}, {Giani}, {Ghinassi}, {Hunt},
  {Maiolino}, {Mannucci}, {Marcucci}, {Sozzi}, {Stefanini}, \&
  {Testi}}]{baffa2001}
{Baffa} C. {et~al.}, 2001, \aap, 378, 722

\bibitem[{{Baraffe} {et~al}\mbox{.}(2003){Baraffe}, {Chabrier}, {Barman},
  {Allard}, \& {Hauschildt}}]{baraffe2003}
{Baraffe} I., {Chabrier} G., {Barman} T.~S., {Allard} F., {Hauschildt} P.~H.,
  2003, \aap, 402, 701

\bibitem[{{Barber} {et~al}\mbox{.}(2006){Barber}, {Tennyson}, {Harris}, \&
  {Tolchenov}}]{barber2006}
{Barber} R.~J., {Tennyson} J., {Harris} G.~J., {Tolchenov} R.~N., 2006, \mnras,
  368, 1087

\bibitem[{{Barry}(1988)}]{barry1988}
{Barry} D.~C., 1988, \apj, 334, 436

\bibitem[{{Bate} {et~al}\mbox{.}(2002){Bate}, {Bonnell}, \& {Bromm}}]{bate2002}
{Bate} M.~R., {Bonnell} I.~A., {Bromm} V., 2002, \mnras, 332, L65

\bibitem[{{Bohlin} \& {Gilliland}(2004)}]{bohlin2004}
{Bohlin} R.~C., {Gilliland} R.~L., 2004, \aj, 127, 3508

\bibitem[{{Borkova} \& {Marsakov}(2004)}]{borkova2004}
{Borkova} T.~V., {Marsakov} V.~A., 2004, Astronomy Letters, 30, 148

\bibitem[{{Borkova} \& {Marsakov}(2005)}]{borkova2005}
{Borkova} T.~V., {Marsakov} V.~A., 2005, Astronomy Reports, 49, 405

\bibitem[{{Borucki} {et~al}\mbox{.}(2011){Borucki}, {Koch}, {Basri}, {Batalha},
  {Brown}, {Bryson}, {Caldwell}, {Christensen-Dalsgaard}, {Cochran}, {DeVore},
  {Dunham}, {Gautier}, {Geary}, {Gilliland}, {Gould}, {Howell}, {Jenkins},
  {Latham}, {Lissauer}, {Marcy}, {Rowe}, {Sasselov}, {Boss}, {Charbonneau},
  {Ciardi}, {Doyle}, {Dupree}, {Ford}, {Fortney}, {Holman}, {Seager},
  {Steffen}, {Tarter}, {Welsh}, {Allen}, {Buchhave}, {Christiansen}, {Clarke},
  {Das}, {D{\'e}sert}, {Endl}, {Fabrycky}, {Fressin}, {Haas}, {Horch},
  {Howard}, {Isaacson}, {Kjeldsen}, {Kolodziejczak}, {Kulesa}, {Li}, {Lucas},
  {Machalek}, {McCarthy}, {MacQueen}, {Meibom}, {Miquel}, {Prsa}, {Quinn},
  {Quintana}, {Ragozzine}, {Sherry}, {Shporer}, {Tenenbaum}, {Torres},
  {Twicken}, {Van Cleve}, {Walkowicz}, {Witteborn}, \& {Still}}]{borucki2011}
{Borucki} W.~J. {et~al.}, 2011, \apj, 736, 19

\bibitem[{{Bouchy} {et~al}\mbox{.}(2011{\natexlab{a}}){Bouchy}, {Bonomo},
  {Santerne}, {Moutou}, {Deleuil}, {D{\'{\i}}az}, {Eggenberger}, {Ehrenreich},
  {Gry}, {Guillot}, {Havel}, {H{\'e}brard}, \& {Udry}}]{bouchy2011b}
{Bouchy} F. {et~al.}, 2011{\natexlab{a}}, \aap, 533, A83

\bibitem[{{Bouchy} {et~al}\mbox{.}(2011{\natexlab{b}}){Bouchy}, {Deleuil},
  {Guillot}, {Aigrain}, {Carone}, {Cochran}, {Almenara}, {Alonso}, {Auvergne},
  {Baglin}, {Barge}, {Bonomo}, {Bord{\'e}}, {Csizmadia}, {de Bondt}, {Deeg},
  {D{\'{\i}}az}, {Dvorak}, {Endl}, {Erikson}, {Ferraz-Mello}, {Fridlund},
  {Gandolfi}, {Gazzano}, {Gibson}, {Gillon}, {Guenther}, {Hatzes}, {Havel},
  {H{\'e}brard}, {Jorda}, {L{\'e}ger}, {Lovis}, {Llebaria}, {Lammer},
  {MacQueen}, {Mazeh}, {Moutou}, {Ofir}, {Ollivier}, {Parviainen},
  {P{\"a}tzold}, {Queloz}, {Rauer}, {Rouan}, {Santerne}, {Schneider},
  {Tingley}, \& {Wuchterl}}]{bouchy2011a}
{Bouchy} F. {et~al.}, 2011{\natexlab{b}}, \aap, 525, A68

\bibitem[{{Burgasser} {et~al}\mbox{.}(2006{\natexlab{a}}){Burgasser},
  {Burrows}, \& {Kirkpatrick}}]{bbk2006}
{Burgasser} A.~J., {Burrows} A., {Kirkpatrick} J.~D., 2006{\natexlab{a}}, \apj,
  639, 1095

\bibitem[{{Burgasser} {et~al}\mbox{.}(2006{\natexlab{b}}){Burgasser},
  {Geballe}, {Leggett}, {Kirkpatrick}, \& {Golimowski}}]{burgasser06}
{Burgasser} A.~J., {Geballe} T.~R., {Leggett} S.~K., {Kirkpatrick} J.~D.,
  {Golimowski} D.~A., 2006{\natexlab{b}}, \apj, 637, 1067

\bibitem[{{Burgasser} {et~al}\mbox{.}(2002){Burgasser}, {Kirkpatrick}, {Brown},
  {Reid}, {Burrows}, {Liebert}, {Matthews}, {Gizis}, {Dahn}, {Monet}, {Cutri},
  \& {Skrutskie}}]{burgasser02}
{Burgasser} A.~J. {et~al.}, 2002, \apj, 564, 421

\bibitem[{{Burgasser} {et~al}\mbox{.}(2000){Burgasser}, {Kirkpatrick}, {Cutri},
  {McCallon}, {Kopan}, {Gizis}, {Liebert}, {Reid}, {Brown}, {Monet}, {Dahn},
  {Beichman}, \& {Skrutskie}}]{burgasser2000}
{Burgasser} A.~J. {et~al.}, 2000, \apjl, 531, L57

\bibitem[{{Burgasser} {et~al}\mbox{.}(2005){Burgasser}, {Kirkpatrick}, \&
  {Lowrance}}]{burgasser2005}
{Burgasser} A.~J., {Kirkpatrick} J.~D., {Lowrance} P.~J., 2005, \aj, 129, 2849

\bibitem[{{Burningham} {et~al}\mbox{.}(2011{\natexlab{a}}){Burningham},
  {Leggett}, {Homeier}, {Saumon}, {Lucas}, {Pinfield}, {Tinney}, {Allard},
  {Marley}, {Jones}, {Murray}, {Ishii}, {Day-Jones}, {Gomes}, \&
  {Zhang}}]{ben2011a}
{Burningham} B. {et~al.}, 2011{\natexlab{a}}, \mnras, 414, 3590

\bibitem[{{Burningham} {et~al}\mbox{.}(2011{\natexlab{b}}){Burningham},
  {Leggett}, {Homeier}, {Saumon}, {Lucas}, {Pinfield}, {Tinney}, {Allard},
  {Marley}, {Jones}, {Murray}, {Ishii}, {Day-Jones}, {Gomes}, \&
  {Zhang}}]{ben11a}
{Burningham} B. {et~al.}, 2011{\natexlab{b}}, \mnras, 414, 3590

\bibitem[{{Burningham} {et~al}\mbox{.}(2010{\natexlab{a}}){Burningham},
  {Leggett}, {Lucas}, {Pinfield}, {Smart}, {Day-Jones}, {Jones}, {Murray},
  {Nickson}, {Tamura}, {Zhang}, {Lodieu}, {Tinney}, \& {Zapatero
  Osorio}}]{ben2010a}
{Burningham} B. {et~al.}, 2010{\natexlab{a}}, \mnras, 404, 1952

\bibitem[{{Burningham} {et~al}\mbox{.}(2011{\natexlab{c}}){Burningham},
  {Lucas}, {Leggett}, {Smart}, {Baker}, {Pinfield}, {Tinney}, {Homeier},
  {Allard}, {Zhang}, {Gomes}, {Day-Jones}, {Jones}, {Kov{\'a}cs}, {Lodieu},
  {Marocco}, {Murray}, \& {Sip{\H o}cz}}]{ben2011b}
{Burningham} B. {et~al.}, 2011{\natexlab{c}}, \mnras, 414, L90

\bibitem[{{Burningham} {et~al}\mbox{.}(2008){Burningham}, {Pinfield},
  {Leggett}, {Tamura}, {Lucas}, {Homeier}, {Day-Jones}, {Jones}, {Clarke},
  {Ishii}, {Kuzuhara}, {Lodieu}, {Zapatero Osorio}, {Venemans}, {Mortlock},
  {Barrado Y Navascu{\'e}s}, {Martin}, \& {Magazz{\`u}}}]{ben2008}
{Burningham} B. {et~al.}, 2008, \mnras, 391, 320

\bibitem[{{Burningham} {et~al}\mbox{.}(2009{\natexlab{a}}){Burningham},
  {Pinfield}, {Leggett}, {Tinney}, {Liu}, {Homeier}, {West}, {Day-Jones},
  {Huelamo}, {Dupuy}, {Zhang}, {Murray}, {Lodieu}, {Barrado Y Navascu{\'e}s},
  {Folkes}, {Galvez-Ortiz}, {Jones}, {Lucas}, {Calderon}, \&
  {Tamura}}]{ben2009}
{Burningham} B. {et~al.}, 2009{\natexlab{a}}, \mnras, 395, 1237

\bibitem[{{Burningham} {et~al}\mbox{.}(2009{\natexlab{b}}){Burningham},
  {Pinfield}, {Leggett}, {Tinney}, {Liu}, {Homeier}, {West}, {Day-Jones},
  {Huelamo}, {Dupuy}, {Zhang}, {Murray}, {Lodieu}, {Barrado Y Navascu{\'e}s},
  {Folkes}, {Galvez-Ortiz}, {Jones}, {Lucas}, {Calderon}, \& {Tamura}}]{ben09}
{Burningham} B. {et~al.}, 2009{\natexlab{b}}, \mnras, 395, 1237

\bibitem[{{Burningham} {et~al}\mbox{.}(2010{\natexlab{b}}){Burningham},
  {Pinfield}, {Lucas}, {Leggett}, {Deacon}, {Tamura}, {Tinney}, {Lodieu},
  {Zhang}, {Huelamo}, {Jones}, {Murray}, {Mortlock}, {Patel}, {Barrado Y
  Navascu{\'e}s}, {Zapatero Osorio}, {Ishii}, {Kuzuhara}, \&
  {Smart}}]{ben2010b}
{Burningham} B. {et~al.}, 2010{\natexlab{b}}, \mnras, 406, 1885

\bibitem[{{Burrows} {et~al}\mbox{.}(2011){Burrows}, {Heng}, \&
  {Nampaisarn}}]{burrows2011}
{Burrows} A., {Heng} K., {Nampaisarn} T., 2011, \apj, 736, 47

\bibitem[{{Carson} {et~al}\mbox{.}(2009){Carson}, {Hiner}, {Villar},
  {Blaschak}, {Rudolph}, \& {Stapelfeldt}}]{carson2009}
{Carson} J.~C., {Hiner} K.~D., {Villar}, III G.~G., {Blaschak} M.~G., {Rudolph}
  A.~L., {Stapelfeldt} K.~R., 2009, \aj, 137, 218

\bibitem[{{Casagrande} {et~al}\mbox{.}(2010){Casagrande}, {Ram{\'{\i}}rez},
  {Mel{\'e}ndez}, {Bessell}, \& {Asplund}}]{casagrande2010}
{Casagrande} L., {Ram{\'{\i}}rez} I., {Mel{\'e}ndez} J., {Bessell} M.,
  {Asplund} M., 2010, \aap, 512, A54

\bibitem[{{Cayrel de Strobel} {et~al}\mbox{.}(2001){Cayrel de Strobel},
  {Soubiran}, \& {Ralite}}]{cayreldestrobel2001}
{Cayrel de Strobel} G., {Soubiran} C., {Ralite} N., 2001, \aap, 373, 159

\bibitem[{{Clarke} {et~al}\mbox{.}(2010){Clarke}, {Pinfield},
  {G{\'a}lvez-Ortiz}, {Jenkins}, {Burningham}, {Deacon}, {Jones}, {Pokorny},
  {Barnes}, \& {Day-Jones}}]{clarke2010}
{Clarke} J.~R.~A. {et~al.}, 2010, \mnras, 402, 575

\bibitem[{{Cooke} \& {Rodgers}(2005)}]{cooke2005}
{Cooke} A., {Rodgers} B., 2005, in Astronomical Society of the Pacific
  Conference Series, Vol. 347, Astronomical Data Analysis Software and Systems
  XIV, {P.~Shopbell, M.~Britton, \& R.~Ebert}, ed., pp. 514--+

\bibitem[{{Cumming} {et~al}\mbox{.}(1999){Cumming}, {Marcy}, \&
  {Butler}}]{cumming1999}
{Cumming} A., {Marcy} G.~W., {Butler} R.~P., 1999, \apj, 526, 890

\bibitem[{{Cushing} {et~al}\mbox{.}(2011){Cushing}, {Kirkpatrick}, {Gelino},
  {Griffith}, {Skrutskie}, {Mainzer}, {Marsh}, {Beichman}, {Burgasser},
  {Prato}, {Simcoe}, {Marley}, {Saumon}, {Freedman}, {Eisenhardt}, \&
  {Wright}}]{cushing2011}
{Cushing} M.~C. {et~al.}, 2011, ArXiv e-prints

\bibitem[{{Day-Jones} {et~al}\mbox{.}(2011){Day-Jones}, {Pinfield}, {Ruiz},
  {Beaumont}, {Burningham}, {Gallardo}, {Gianninas}, {Bergeron}, {Napiwotzki},
  {Jenkins}, {Zhang}, {Murray}, {Catal{\'a}n}, \& {Gomes}}]{avril2011}
{Day-Jones} A.~C. {et~al.}, 2011, \mnras, 410, 705

\bibitem[{{Deleuil} {et~al}\mbox{.}(2008){Deleuil}, {Deeg}, {Alonso}, {Bouchy},
  {Rouan}, {Auvergne}, {Baglin}, {Aigrain}, {Almenara}, {Barbieri}, {Barge},
  {Bruntt}, {Bord{\'e}}, {Collier Cameron}, {Csizmadia}, {de La Reza},
  {Dvorak}, {Erikson}, {Fridlund}, {Gandolfi}, {Gillon}, {Guenther}, {Guillot},
  {Hatzes}, {H{\'e}brard}, {Jorda}, {Lammer}, {L{\'e}ger}, {Llebaria},
  {Loeillet}, {Mayor}, {Mazeh}, {Moutou}, {Ollivier}, {P{\"a}tzold}, {Pont},
  {Queloz}, {Rauer}, {Schneider}, {Shporer}, {Wuchterl}, \&
  {Zucker}}]{deleuil2008}
{Deleuil} M. {et~al.}, 2008, \aap, 491, 889

\bibitem[{{Delorme} {et~al}\mbox{.}(2008){Delorme}, {Delfosse}, {Albert},
  {Artigau}, {Forveille}, {Reyl{\'e}}, {Allard}, {Homeier}, {Robin}, {Willott},
  {Liu}, \& {Dupuy}}]{delorme2008}
{Delorme} P. {et~al.}, 2008, \aap, 482, 961

\bibitem[{{do Nascimento} {et~al}\mbox{.}(2010){do Nascimento}, {da Costa}, \&
  {de Medeiros}}]{donascimento2010}
{do Nascimento} J.~D., {da Costa} J.~S., {de Medeiros} J.~R., 2010, \aap, 519,
  A101

\bibitem[{{Elias} {et~al}\mbox{.}(2006){Elias}, {Joyce}, {Liang}, {Muller},
  {Hileman}, \& {George}}]{elias2006}
{Elias} J.~H., {Joyce} R.~R., {Liang} M., {Muller} G.~P., {Hileman} E.~A.,
  {George} J.~R., 2006, in Presented at the Society of Photo-Optical
  Instrumentation Engineers (SPIE) Conference, Vol. 6269, Society of
  Photo-Optical Instrumentation Engineers (SPIE) Conference Series

\bibitem[{{Endl} {et~al}\mbox{.}(2002){Endl}, {K{\"u}rster}, {Els}, {Hatzes},
  {Cochran}, {Dennerl}, \& {D{\"o}bereiner}}]{endl2002}
{Endl} M., {K{\"u}rster} M., {Els} S., {Hatzes} A.~P., {Cochran} W.~D.,
  {Dennerl} K., {D{\"o}bereiner} S., 2002, \aap, 392, 671

\bibitem[{{Epchtein} {et~al}\mbox{.}(1997){Epchtein}, {de Batz}, {Capoani},
  {Chevallier}, {Copet}, {Fouqu{\'e}}, {Lacombe}, {Le Bertre}, {Pau}, {Rouan},
  {Ruphy}, {Simon}, {Tiph{\`e}ne}, {Burton}, {Bertin}, {Deul}, {Habing},
  {Borsenberger}, {Dennefeld}, {Guglielmo}, {Loup}, {Mamon}, {Ng}, {Omont},
  {Provost}, {Renault}, {Tanguy}, {Kimeswenger}, {Kienel}, {Garzon}, {Persi},
  {Ferrari-Toniolo}, {Robin}, {Paturel}, {Vauglin}, {Forveille}, {Delfosse},
  {Hron}, {Schultheis}, {Appenzeller}, {Wagner}, {Balazs}, {Holl},
  {L{\'e}pine}, {Boscolo}, {Picazzio}, {Duc}, \& {Mennessier}}]{epchtein1997}
{Epchtein} N. {et~al.}, 1997, The Messenger, 87, 27

\bibitem[{{Faherty} {et~al}\mbox{.}(2010){Faherty}, {Burgasser}, {West},
  {Bochanski}, {Cruz}, {Shara}, \& {Walter}}]{faherty2010}
{Faherty} J.~K., {Burgasser} A.~J., {West} A.~A., {Bochanski} J.~J., {Cruz}
  K.~L., {Shara} M.~M., {Walter} F.~M., 2010, \aj, 139, 176

\bibitem[{{Freytag} {et~al}\mbox{.}(2010){Freytag}, {Allard}, {Ludwig},
  {Homeier}, \& {Steffen}}]{bdco5bold}
{Freytag} B., {Allard} F., {Ludwig} H., {Homeier} D., {Steffen} M., 2010, \aap,
  513, A19+

\bibitem[{{G{\'a}lvez-Ortiz} {et~al}\mbox{.}(2010){G{\'a}lvez-Ortiz}, {Clarke},
  {Pinfield}, {Jenkins}, {Folkes}, {P{\'e}rez}, {Day-Jones}, {Burningham},
  {Jones}, {Barnes}, \& {Pokorny}}]{maricruz2010}
{G{\'a}lvez-Ortiz} M.~C. {et~al.}, 2010, \mnras, 409, 552

\bibitem[{{Giridhar} \& {Goswami}(2002)}]{giridhar2002}
{Giridhar} S., {Goswami} A., 2002, Bulletin of the Astronomical Society of
  India, 30, 501

\bibitem[{{Gliese} \& {Jahreiss}(1991)}]{gliese1991}
{Gliese} W., {Jahreiss} H., 1991, NASA STI/Recon Technical Report A, 923, 33932

\bibitem[{{Gontcharov}(2006)}]{gontcharov2006}
{Gontcharov} G.~A., 2006, Astronomy Letters, 32, 759

\bibitem[{{Goodwin} \& {Whitworth}(2007)}]{goodwin2007}
{Goodwin} S.~P., {Whitworth} A., 2007, \aap, 466, 943

\bibitem[{{Halbwachs} {et~al}\mbox{.}(2000){Halbwachs}, {Arenou}, {Mayor},
  {Udry}, \& {Queloz}}]{halbwachs2000}
{Halbwachs} J.~L., {Arenou} F., {Mayor} M., {Udry} S., {Queloz} D., 2000, \aap,
  355, 581

\bibitem[{{Halbwachs} {et~al}\mbox{.}(2003){Halbwachs}, {Mayor}, {Udry}, \&
  {Arenou}}]{halbwachs2003}
{Halbwachs} J.~L., {Mayor} M., {Udry} S., {Arenou} F., 2003, \aap, 397, 159

\bibitem[{{Hall} {et~al}\mbox{.}(2007){Hall}, {Lockwood}, \&
  {Skiff}}]{hall2007}
{Hall} J.~C., {Lockwood} G.~W., {Skiff} B.~A., 2007, \aj, 133, 862

\bibitem[{{Haywood}(2001)}]{haywood2001}
{Haywood} M., 2001, \mnras, 325, 1365

\bibitem[{{Heiter} \& {Luck}(2003)}]{heiter2003}
{Heiter} U., {Luck} R.~E., 2003, \aj, 126, 2015

\bibitem[{{Hewett} {et~al}\mbox{.}(2006){Hewett}, {Warren}, {Leggett}, \&
  {Hodgkin}}]{Hewett06}
{Hewett} P.~C., {Warren} S.~J., {Leggett} S.~K., {Hodgkin} S.~T., 2006, \mnras,
  367, 454

\bibitem[{{Holmberg} {et~al}\mbox{.}(2007){Holmberg}, {Nordstr{\"o}m}, \&
  {Andersen}}]{holmberg2007}
{Holmberg} J., {Nordstr{\"o}m} B., {Andersen} J., 2007, \aap, 475, 519

\bibitem[{{Holmberg} {et~al}\mbox{.}(2009){Holmberg}, {Nordstr{\"o}m}, \&
  {Andersen}}]{holmberg2009}
{Holmberg} J., {Nordstr{\"o}m} B., {Andersen} J., 2009, \aap, 501, 941

\bibitem[{{Ibukiyama} \& {Arimoto}(2002)}]{ibukiyama2002}
{Ibukiyama} A., {Arimoto} N., 2002, \aap, 394, 927

\bibitem[{{Johnson} {et~al}\mbox{.}(2011){Johnson}, {Apps}, {Gazak}, {Crepp},
  {Crossfield}, {Howard}, {Marcy}, {Morton}, {Chubak}, \&
  {Isaacson}}]{johnson2011}
{Johnson} J.~A. {et~al.}, 2011, \apj, 730, 79

\bibitem[{{Karata{\c s}} {et~al}\mbox{.}(2005){Karata{\c s}}, {Bilir}, \&
  {Schuster}}]{karatas2005}
{Karata{\c s}} Y., {Bilir} S., {Schuster} W.~J., 2005, \mnras, 360, 1345

\bibitem[{{Kirkpatrick}(2005)}]{kirkpatrick2005}
{Kirkpatrick} J.~D., 2005, \araa, 43, 195

\bibitem[{{Kirkpatrick} {et~al}\mbox{.}(2011){Kirkpatrick}, {Cushing},
  {Gelino}, {Griffith}, {Skrutskie}, {Marsh}, {Wright}, {Mainzer},
  {Eisenhardt}, {McLean}, {Thompson}, {Bauer}, {Benford}, {Bridge}, {Lake},
  {Petty}, {Stanford}, {Tsai}, {Bailey}, {Beichman}, {Bochanski}, {Burgasser},
  {Capak}, {Cruz}, {Hinz}, {Kartaltepe}, {Knox}, {Manohar}, {Masters},
  {Morales-Calderon}, {Prato}, {Rodigas}, {Salvato}, {Schurr}, {Scoville},
  {Simcoe}, {Stapelfeldt}, {Stern}, {Stock}, \& {Vacca}}]{kirkpatrick2011}
{Kirkpatrick} J.~D. {et~al.}, 2011, ArXiv e-prints

\bibitem[{{Knapp} {et~al}\mbox{.}(2004){Knapp}, {Leggett}, {Fan}, {Marley},
  {Geballe}, {Golimowski}, {Finkbeiner}, {Gunn}, {Hennawi}, {Ivezi{\'c}},
  {Lupton}, {Schlegel}, {Strauss}, {Tsvetanov}, {Chiu}, {Hoversten},
  {Glazebrook}, {Zheng}, {Hendrickson}, {Williams}, {Uomoto}, {Vrba}, {Henden},
  {Luginbuhl}, {Guetter}, {Munn}, {Canzian}, {Schneider}, \&
  {Brinkmann}}]{knapp04}
{Knapp} G.~R. {et~al.}, 2004, \aj, 127, 3553

\bibitem[{{Kovtyukh} {et~al}\mbox{.}(2003){Kovtyukh}, {Soubiran}, {Belik}, \&
  {Gorlova}}]{kovtyukh2003}
{Kovtyukh} V.~V., {Soubiran} C., {Belik} S.~I., {Gorlova} N.~I., 2003, \aap,
  411, 559

\bibitem[{{Latham} {et~al}\mbox{.}(2002){Latham}, {Stefanik}, {Torres},
  {Davis}, {Mazeh}, {Carney}, {Laird}, \& {Morse}}]{latham2002}
{Latham} D.~W., {Stefanik} R.~P., {Torres} G., {Davis} R.~J., {Mazeh} T.,
  {Carney} B.~W., {Laird} J.~B., {Morse} J.~A., 2002, \aj, 124, 1144

\bibitem[{{Lawler} {et~al}\mbox{.}(2009){Lawler}, {Beichman}, {Bryden},
  {Ciardi}, {Tanner}, {Su}, {Stapelfeldt}, {Lisse}, \& {Harker}}]{lawler2009}
{Lawler} S.~M. {et~al.}, 2009, \apj, 705, 89

\bibitem[{{Lawrence} {et~al}\mbox{.}(2007){Lawrence}, {Warren}, {Almaini},
  {Edge}, {Hambly}, {Jameson}, {Lucas}, {Casali}, {Adamson}, {Dye}, {Emerson},
  {Foucaud}, {Hewett}, {Hirst}, {Hodgkin}, {Irwin}, {Lodieu}, {McMahon},
  {Simpson}, {Smail}, {Mortlock}, \& {Folger}}]{lawrence2007}
{Lawrence} A. {et~al.}, 2007, \mnras, 379, 1599

\bibitem[{{Lebreton} {et~al}\mbox{.}(1999){Lebreton}, {Perrin}, {Cayrel},
  {Baglin}, \& {Fernandes}}]{lebreton1999}
{Lebreton} Y., {Perrin} M.-N., {Cayrel} R., {Baglin} A., {Fernandes} J., 1999,
  \aap, 350, 587

\bibitem[{{Leggett} {et~al}\mbox{.}(2010{\natexlab{a}}){Leggett}, {Burningham},
  {Saumon}, {Marley}, {Warren}, {Smart}, {Jones}, {Lucas}, {Pinfield}, \&
  {Tamura}}]{sandy2010a}
{Leggett} S.~K. {et~al.}, 2010{\natexlab{a}}, \apj, 710, 1627

\bibitem[{{Leggett} {et~al}\mbox{.}(2010{\natexlab{b}}){Leggett}, {Burningham},
  {Saumon}, {Marley}, {Warren}, {Smart}, {Jones}, {Lucas}, {Pinfield}, \&
  {Tamura}}]{sandy10}
{Leggett} S.~K. {et~al.}, 2010{\natexlab{b}}, \apj, 710, 1627

\bibitem[{{Leggett} {et~al}\mbox{.}(2006){Leggett}, {Currie}, {Varricatt},
  {Hawarden}, {Adamson}, {Buckle}, {Carroll}, {Davies}, {Davis}, {Kerr},
  {Kuhn}, {Seigar}, \& {Wold}}]{sandy2006}
{Leggett} S.~K. {et~al.}, 2006, \mnras, 373, 781

\bibitem[{{Leggett} {et~al}\mbox{.}(2010{\natexlab{c}}){Leggett}, {Saumon},
  {Burningham}, {Cushing}, {Marley}, \& {Pinfield}}]{sandy2010b}
{Leggett} S.~K., {Saumon} D., {Burningham} B., {Cushing} M.~C., {Marley} M.~S.,
  {Pinfield} D.~J., 2010{\natexlab{c}}, \apj, 720, 252

\bibitem[{{Liu} {et~al}\mbox{.}(2007){Liu}, {Leggett}, \& {Chiu}}]{liu07}
{Liu} M.~C., {Leggett} S.~K., {Chiu} K., 2007, \apj, 660, 1507

\bibitem[{{Lodieu} {et~al}\mbox{.}(2007){Lodieu}, {Pinfield}, {Leggett},
  {Jameson}, {Mortlock}, {Warren}, {Burningham}, {Lucas}, {Chiu}, {Liu},
  {Venemans}, {McMahon}, {Allard}, {Baraffe}, {Barrado y Navascu{\'e}s},
  {Carraro}, {Casewell}, {Chabrier}, {Chappelle}, {Clarke}, {Day-Jones},
  {Deacon}, {Dobbie}, {Folkes}, {Hambly}, {Hewett}, {Hodgkin}, {Jones},
  {Kendall}, {Magazz{\`u}}, {Mart{\'{\i}}n}, {McCaughrean}, {Nakajima},
  {Pavlenko}, {Tamura}, {Tinney}, \& {Zapatero Osorio}}]{lodieu2007}
{Lodieu} N. {et~al.}, 2007, \mnras, 379, 1423

\bibitem[{{Lucas} {et~al}\mbox{.}(2010){Lucas}, {Tinney}, {Burningham},
  {Leggett}, {Pinfield}, {Smart}, {Jones}, {Marocco}, {Barber}, {Yurchenko},
  {Tennyson}, {Ishii}, {Tamura}, {Day-Jones}, {Adamson}, {Allard}, \&
  {Homeier}}]{phil2010}
{Lucas} P.~W. {et~al.}, 2010, \mnras, 408, L56

\bibitem[{{Luck} \& {Heiter}(2006)}]{luck2006}
{Luck} R.~E., {Heiter} U., 2006, \aj, 131, 3069

\bibitem[{{Mainzer} {et~al}\mbox{.}(2011){Mainzer}, {Cushing}, {Skrutskie},
  {Gelino}, {Kirkpatrick}, {Jarrett}, {Masci}, {Marley}, {Saumon}, {Wright},
  {Beaton}, {Dietrich}, {Eisenhardt}, {Garnavich}, {Kuhn}, {Leisawitz},
  {Marsh}, {McLean}, {Padgett}, \& {Rueff}}]{mainzer2011}
{Mainzer} A. {et~al.}, 2011, \apj, 726, 30

\bibitem[{{Mashonkina} \& {Gehren}(2001)}]{mashonkina2001}
{Mashonkina} L., {Gehren} T., 2001, \aap, 376, 232

\bibitem[{{Mashonkina} {et~al}\mbox{.}(2007){Mashonkina}, {Korn}, \&
  {Przybilla}}]{mashonkina2007}
{Mashonkina} L., {Korn} A.~J., {Przybilla} N., 2007, \aap, 461, 261

\bibitem[{{Mishenina} {et~al}\mbox{.}(2004){Mishenina}, {Soubiran}, {Kovtyukh},
  \& {Korotin}}]{mishenina2004}
{Mishenina} T.~V., {Soubiran} C., {Kovtyukh} V.~V., {Korotin} S.~A., 2004,
  \aap, 418, 551

\bibitem[{{Munn} {et~al}\mbox{.}(2004){Munn}, {Monet}, {Levine}, {Canzian},
  {Pier}, {Harris}, {Lupton}, {Ivezi{\'c}}, {Hindsley}, {Hennessy},
  {Schneider}, \& {Brinkmann}}]{Munn2004}
{Munn} J.~A. {et~al.}, 2004, \aj, 127, 3034

\bibitem[{{Murray} {et~al}\mbox{.}(2011){Murray}, {Burningham}, {Jones},
  {Pinfield}, {Lucas}, {Leggett}, {Tinney}, {Day-Jones}, {Weights}, {Lodieu},
  {P{\'e}rez Prieto}, {Nickson}, {Zhang}, {Clarke}, {Jenkins}, \&
  {Tamura}}]{dm2011}
{Murray} D.~N. {et~al.}, 2011, \mnras, 414, 575

\bibitem[{{Nidever} {et~al}\mbox{.}(2002){Nidever}, {Marcy}, {Butler},
  {Fischer}, \& {Vogt}}]{nidever2002}
{Nidever} D.~L., {Marcy} G.~W., {Butler} R.~P., {Fischer} D.~A., {Vogt} S.~S.,
  2002, \apjs, 141, 503

\bibitem[{{Nordstr{\"o}m} {et~al}\mbox{.}(2004){Nordstr{\"o}m}, {Mayor},
  {Andersen}, {Holmberg}, {Pont}, {J{\o}rgensen}, {Olsen}, {Udry}, \&
  {Mowlavi}}]{nordstroem2004}
{Nordstr{\"o}m} B. {et~al.}, 2004, \aap, 418, 989

\bibitem[{{Pinfield} {et~al}\mbox{.}(2008){Pinfield}, {Burningham}, {Tamura},
  {Leggett}, {Lodieu}, {Lucas}, {Mortlock}, {Warren}, {Homeier}, {Ishii},
  {Deacon}, {McMahon}, {Hewett}, {Osori}, {Martin}, {Jones}, {Venemans},
  {Day-Jones}, {Dobbie}, {Folkes}, {Dye}, {Allard}, {Baraffe}, {Barrado Y
  Navascu{\'e}s}, {Casewell}, {Chiu}, {Chabrier}, {Clarke}, {Hodgkin},
  {Magazz{\`u}}, {McCaughrean}, {Nakajima}, {Pavlenko}, \& {Tinney}}]{me2008}
{Pinfield} D.~J. {et~al.}, 2008, \mnras, 390, 304

\bibitem[{{Pinfield} {et~al}\mbox{.}(2006){Pinfield}, {Jones}, {Lucas},
  {Kendall}, {Folkes}, {Day-Jones}, {Chappelle}, \& {Steele}}]{me2006}
{Pinfield} D.~J., {Jones} H.~R.~A., {Lucas} P.~W., {Kendall} T.~R., {Folkes}
  S.~L., {Day-Jones} A.~C., {Chappelle} R.~J., {Steele} I.~A., 2006, \mnras,
  368, 1281

\bibitem[{{Pinfield} {et~al}\mbox{.}(2005){Pinfield}, {Jones}, \&
  {Steele}}]{me2005}
{Pinfield} D.~J., {Jones} H.~R.~A., {Steele} I.~A., 2005, \pasp, 117, 173

\bibitem[{{Pont} {et~al}\mbox{.}(2005{\natexlab{a}}){Pont}, {Bouchy}, {Melo},
  {Santos}, {Mayor}, {Queloz}, \& {Udry}}]{pont2005b}
{Pont} F., {Bouchy} F., {Melo} C., {Santos} N.~C., {Mayor} M., {Queloz} D.,
  {Udry} S., 2005{\natexlab{a}}, \aap, 438, 1123

\bibitem[{{Pont} {et~al}\mbox{.}(2005{\natexlab{b}}){Pont}, {Melo}, {Bouchy},
  {Udry}, {Queloz}, {Mayor}, \& {Santos}}]{pont2005a}
{Pont} F., {Melo} C.~H.~F., {Bouchy} F., {Udry} S., {Queloz} D., {Mayor} M.,
  {Santos} N.~C., 2005{\natexlab{b}}, \aap, 433, L21

\bibitem[{{Ram{\'{\i}}rez} {et~al}\mbox{.}(2007){Ram{\'{\i}}rez}, {Allende
  Prieto}, \& {Lambert}}]{ramirez2007}
{Ram{\'{\i}}rez} I., {Allende Prieto} C., {Lambert} D.~L., 2007, \aap, 465, 271

\bibitem[{{Reid} {et~al}\mbox{.}(2007){Reid}, {Cruz}, \& {Allen}}]{reid2007}
{Reid} I.~N., {Cruz} K.~L., {Allen} P.~R., 2007, \aj, 133, 2825

\bibitem[{{Reyl{\'e}} {et~al}\mbox{.}(2010){Reyl{\'e}}, {Delorme}, {Willott},
  {Albert}, {Delfosse}, {Forveille}, {Artigau}, {Malo}, {Hill}, \&
  {Doyon}}]{reyle2010}
{Reyl{\'e}} C. {et~al.}, 2010, \aap, 522, A112

\bibitem[{{Rocha-Pinto} \& {Maciel}(1998)}]{rochapinto1998}
{Rocha-Pinto} H.~J., {Maciel} W.~J., 1998, \mnras, 298, 332

\bibitem[{{Saumon} {et~al}\mbox{.}(1994){Saumon}, {Bergeron}, {Lunine},
  {Hubbard}, \& {Burrows}}]{saumon94}
{Saumon} D., {Bergeron} P., {Lunine} J.~I., {Hubbard} W.~B., {Burrows} A.,
  1994, \apj, 424, 333

\bibitem[{{Saumon} {et~al}\mbox{.}(2007){Saumon}, {Marley}, {Leggett},
  {Geballe}, {Stephens}, {Golimowski}, {Cushing}, {Fan}, {Rayner}, {Lodders},
  \& {Freedman}}]{saumon2007}
{Saumon} D. {et~al.}, 2007, \apj, 656, 1136

\bibitem[{{Scholz}(2010)}]{scholz2010}
{Scholz} R.-D., 2010, \aap, 515, A92

\bibitem[{{Shi} {et~al}\mbox{.}(2004){Shi}, {Gehren}, \& {Zhao}}]{shi2004}
{Shi} J.~R., {Gehren} T., {Zhao} G., 2004, \aap, 423, 683

\bibitem[{{Skrutskie} {et~al}\mbox{.}(2006){Skrutskie}, {Cutri}, {Stiening},
  {Weinberg}, {Schneider}, {Carpenter}, {Beichman}, {Capps}, {Chester},
  {Elias}, {Huchra}, {Liebert}, {Lonsdale}, {Monet}, {Price}, {Seitzer},
  {Jarrett}, {Kirkpatrick}, {Gizis}, {Howard}, {Evans}, {Fowler}, {Fullmer},
  {Hurt}, {Light}, {Kopan}, {Marsh}, {McCallon}, {Tam}, {Van Dyk}, \&
  {Wheelock}}]{skrutskie2006}
{Skrutskie} M.~F. {et~al.}, 2006, \aj, 131, 1163

\bibitem[{{Stamatellos} {et~al}\mbox{.}(2007){Stamatellos}, {Hubber}, \&
  {Whitworth}}]{stamatellos2007}
{Stamatellos} D., {Hubber} D.~A., {Whitworth} A.~P., 2007, \mnras, 382, L30

\bibitem[{{Sumi} {et~al}\mbox{.}(2011){Sumi}, {Kamiya}, {Bennett}, {Bond},
  {Abe}, {Botzler}, {Fukui}, {Furusawa}, {Hearnshaw}, {Itow}, {Kilmartin},
  {Korpela}, {Lin}, {Ling}, {Masuda}, {Matsubara}, {Miyake}, {Motomura},
  {Muraki}, {Nagaya}, {Nakamura}, {Ohnishi}, {Okumura}, {Perrott},
  {Rattenbury}, {Saito}, {Sako}, {Sullivan}, {Sweatman}, {Tristram}, {Udalski},
  {Szyma{\'n}ski}, {Kubiak}, {Pietrzy{\'n}ski}, {Poleski}, {Soszy{\'n}ski},
  {Wyrzykowski}, {Ulaczyk}, \& {Microlensing Observations in Astrophysics (MOA)
  Collaboration}}]{sumi2011}
{Sumi} T. {et~al.}, 2011, \nat, 473, 349

\bibitem[{{Takeda} {et~al}\mbox{.}(2007{\natexlab{a}}){Takeda}, {Ford},
  {Sills}, {Rasio}, {Fischer}, \& {Valenti}}]{takeda2007a}
{Takeda} G., {Ford} E.~B., {Sills} A., {Rasio} F.~A., {Fischer} D.~A.,
  {Valenti} J.~A., 2007{\natexlab{a}}, \apjs, 168, 297

\bibitem[{{Takeda} {et~al}\mbox{.}(2010){Takeda}, {Honda}, {Kawanomoto},
  {Ando}, \& {Sakurai}}]{takeda2010}
{Takeda} Y., {Honda} S., {Kawanomoto} S., {Ando} H., {Sakurai} T., 2010, \aap,
  515, A93

\bibitem[{{Takeda} {et~al}\mbox{.}(2007{\natexlab{b}}){Takeda}, {Kawanomoto},
  {Honda}, {Ando}, \& {Sakurai}}]{takeda2007b}
{Takeda} Y., {Kawanomoto} S., {Honda} S., {Ando} H., {Sakurai} T.,
  2007{\natexlab{b}}, \aap, 468, 663

\bibitem[{{Valenti} \& {Fischer}(2005)}]{valenti2005}
{Valenti} J.~A., {Fischer} D.~A., 2005, \apjs, 159, 141

\bibitem[{{van Leeuwen}(2007)}]{vanleeuwen2007}
{van Leeuwen} F., 2007, \aap, 474, 653

\bibitem[{{Warren} {et~al}\mbox{.}(2007){Warren}, {Mortlock}, {Leggett},
  {Pinfield}, {Homeier}, {Dye}, {Jameson}, {Lodieu}, {Lucas}, {Adamson},
  {Allard}, {Barrado Y Navascu{\'e}s}, {Casali}, {Chiu}, {Hambly}, {Hewett},
  {Hirst}, {Irwin}, {Lawrence}, {Liu}, {Mart{\'{\i}}n}, {Smart}, {Valdivielso},
  \& {Venemans}}]{warren2007}
{Warren} S.~J. {et~al.}, 2007, \mnras, 381, 1400

\bibitem[{{Wright} {et~al}\mbox{.}(2010){Wright}, {Eisenhardt}, {Mainzer},
  {Ressler}, {Cutri}, {Jarrett}, {Kirkpatrick}, {Padgett}, {McMillan},
  {Skrutskie}, {Stanford}, {Cohen}, {Walker}, {Mather}, {Leisawitz}, {Gautier},
  {McLean}, {Benford}, {Lonsdale}, {Blain}, {Mendez}, {Irace}, {Duval}, {Liu},
  {Royer}, {Heinrichsen}, {Howard}, {Shannon}, {Kendall}, {Walsh}, {Larsen},
  {Cardon}, {Schick}, {Schwalm}, {Abid}, {Fabinsky}, {Naes}, \&
  {Tsai}}]{wright2010}
{Wright} E.~L. {et~al.}, 2010, \aj, 140, 1868

\bibitem[{{Wright} {et~al}\mbox{.}(2004){Wright}, {Marcy}, {Butler}, \&
  {Vogt}}]{wright2004}
{Wright} J.~T., {Marcy} G.~W., {Butler} R.~P., {Vogt} S.~S., 2004, \apjs, 152,
  261

\bibitem[{{Wu} {et~al}\mbox{.}(2011){Wu}, {Singh}, {Prugniel}, {Gupta}, \&
  {Koleva}}]{wu2011}
{Wu} Y., {Singh} H.~P., {Prugniel} P., {Gupta} R., {Koleva} M., 2011, \aap,
  525, A71

\bibitem[{{York} {et~al}\mbox{.}(2000){York}, {Adelman}, {Anderson},
  {Anderson}, {Annis}, {Bahcall}, {Bakken}, {Barkhouser}, {Bastian}, {Berman},
  {Boroski}, {Bracker}, {Briegel}, {Briggs}, {Brinkmann}, {Brunner}, {Burles},
  {Carey}, {Carr}, {Castander}, {Chen}, {Colestock}, {Connolly}, {Crocker},
  {Csabai}, {Czarapata}, {Davis}, {Doi}, {Dombeck}, {Eisenstein}, {Ellman},
  {Elms}, {Evans}, {Fan}, {Federwitz}, {Fiscelli}, {Friedman}, {Frieman},
  {Fukugita}, {Gillespie}, {Gunn}, {Gurbani}, {de Haas}, {Haldeman}, {Harris},
  {Hayes}, {Heckman}, {Hennessy}, {Hindsley}, {Holm}, {Holmgren}, {Huang},
  {Hull}, {Husby}, {Ichikawa}, {Ichikawa}, {Ivezi{\'c}}, {Kent}, {Kim},
  {Kinney}, {Klaene}, {Kleinman}, {Kleinman}, {Knapp}, {Korienek}, {Kron},
  {Kunszt}, {Lamb}, {Lee}, {Leger}, {Limmongkol}, {Lindenmeyer}, {Long},
  {Loomis}, {Loveday}, {Lucinio}, {Lupton}, {MacKinnon}, {Mannery}, {Mantsch},
  {Margon}, {McGehee}, {McKay}, {Meiksin}, {Merelli}, {Monet}, {Munn},
  {Narayanan}, {Nash}, {Neilsen}, {Neswold}, {Newberg}, {Nichol}, {Nicinski},
  {Nonino}, {Okada}, {Okamura}, {Ostriker}, {Owen}, {Pauls}, {Peoples},
  {Peterson}, {Petravick}, {Pier}, {Pope}, {Pordes}, {Prosapio},
  {Rechenmacher}, {Quinn}, {Richards}, {Richmond}, {Rivetta}, {Rockosi},
  {Ruthmansdorfer}, {Sandford}, {Schlegel}, {Schneider}, {Sekiguchi}, {Sergey},
  {Shimasaku}, {Siegmund}, {Smee}, {Smith}, {Snedden}, {Stone}, {Stoughton},
  {Strauss}, {Stubbs}, {SubbaRao}, {Szalay}, {Szapudi}, {Szokoly}, {Thakar},
  {Tremonti}, {Tucker}, {Uomoto}, {Vanden Berk}, {Vogeley}, {Waddell}, {Wang},
  {Watanabe}, {Weinberg}, {Yanny}, {Yasuda}, \& {SDSS
  Collaboration}}]{York2000}
{York} D.~G. {et~al.}, 2000, \aj, 120, 1579

\bibitem[{{Zhang} {et~al}\mbox{.}(2010){Zhang}, {Pinfield}, {Day-Jones},
  {Burningham}, {Jones}, {Yu}, {Jenkins}, {Han}, {G{\'a}lvez-Ortiz},
  {Gallardo}, {Garc{\'{\i}}a-P{\'e}rez}, {Weights}, {Tinney}, \&
  {Pokorny}}]{zenghua2010}
{Zhang} Z.~H. {et~al.}, 2010, \mnras, 404, 1817

\end{thebibliography}

\end{document}